\documentclass[article]{elsarticle}
\usepackage{float,graphicx,amsmath,amsthm,amssymb,subcaption,caption,bbold,hyperref}
\newcommand{\bx}{\mathbf{x}}
\newcommand{\be}{\mathbf{e}}
\newcommand{\ba}{\mathbf{a}}
\newcommand{\bc}{\mathbf{c}}

\newcommand{\bd}{\mathbf{d}}
\newcommand{\bu}{\mathbf{u}}
\newcommand{\dd}{\mathrm{d}}
\newcommand{\bphi}{\boldsymbol{\Phi}}
\newcommand{\bvphi}{\boldsymbol{\varphi}}

\newcommand{\bdelta}{\boldsymbol{\delta}}
\journal{Journal of Computational and Applied Mathematics }
\begin{document}
\begin{frontmatter}
	\title{Computing cost-effective particle trajectories in numerically calculated incompressible fluids using geometric methods.}

\author{Benjamin K Tapley}
\ead{benjamin.tapley@ntnu.no}
\address{Department of Mathematical Sciences, The Norwegian University of Science and Technology, 7491 Trondheim, Norway}

	\begin{abstract}
	We present an novel algorithm for tracking massless solid particles in a divergence-free velocity field that is only available at discrete points in space and time such as those arising from a direct numerical simulation of Navier-Stokes. The algorithm creates a divergence-free approximation to the numerical field using matrix valued radial basis functions, which is integrated in time using a volume-preserving map. The resulting method is able to calculate accurate trajectories in a helical vortex using much larger step-sizes and a far lower number of interpolation points which results in a more efficient algorithm compared to a conventional scheme.
\end{abstract}
	
	\begin{keyword}
		Computational fluid dynamics \sep Tracer particles \sep Semi-Lagrangian \sep Radial basis functions \sep Geometric integration 
	\end{keyword}
	
\end{frontmatter}


	\section{Introduction}
	In this paper we are solving for the trajectory $\bx=\bx(t)$ of massless solid particles (also called \textit{tracer particles}) in a vector field that is defined on discrete points in space and time, for example, the result of a direct numerical simulation (DNS) of the Navier-Stokes equation. When $\bx(t)$ is parameterized by the same variable $t$ as the underlying fluid field, then $\bx(t)$ is also referred to as a \textit{pathline}, which is found by solving the ODE
	\begin{equation}\label{ODE}
	\frac{\dd \bx}{\dd t} = \bu(\bx,t),
	\end{equation}
	where $\bu(\bx,t):\mathbb{R}^3\times\mathbb{R}\rightarrow\mathbb{R}^3$ is assumed to be a sufficiently regular vector field that satisfies $\nabla\cdot\bu = 0$. Here, sufficiently regular means that the error associated with the DNS is small in comparison to the accuracy requirements of the resulting particle pathlines. One of the challenges here is that $\bu$ is not a known function of $\bx$ and $t$, but is only available on a discrete set of points in space and time, usually on a regularly spaced grid, hence one must employ an interpolation scheme before equation \eqref{ODE} is suitable for numerical treatment. Such simulations are carried out in order to derive statistical results that determine macroscopic fluid properties, see for example \cite{review} for a review of Lagrangian particle tracking. In addition, the calculation of \textit{streamlines}, which are found by solving equation \eqref{ODE} with a frozen velocity field is important in Lagrangian advection schemes. Application dictates that in order to gain statistically reliable results, one must conduct simulations with millions of particles, which requires efficiently calculating pathlines that obey the qualitative behaviour of the exact solution. The main geometric property of fluid fields are that they are divergence-free and hence an accurate solution to equation \eqref{ODE} should preserve volumes along pathlines of $\bu(\bx,t)$. The problem of solving equation \eqref{ODE} is typically divided into two steps: 
	\begin{itemize}
		\item[1.] Finding a spatial approximation to the discrete fluid velocity field using a 3D vector interpolation scheme 
		\begin{equation}
			\bar{\bu}(\bx,t_0)\approx \bu(\bx,t_0),
		\end{equation}
		where $\bar{\bu}(\bx,t_0)$ is the interpolating vector field that is a known, continuous function of space (but not necessarily time) and is calculated from a set of data points $\{\bx_i,\bu(\bx_i,t_0)\}_{i=1}^N$ from $N$ nearby grid points $\bx_i$ at some fixed time $t_0$.
		\item[2.] Integrating the resulting ODE
		\begin{equation}
			\frac{\dd }{\dd t}\bx(t) = \bar{\bu}(\bx,t),
		\end{equation}
		which is done using a numerical method to find an approximation to $\bx(t+h)$.
	\end{itemize}
	One of the goals of this paper is to propose the use of radial basis function (RBF) interpolation as an approach to address interpolation of numerically calculated fluid fields. Such situations go beyond the calculation of pathlines and can include particle advection schemes, Lagrangian methods for numerical turbulence, tracking inertial and/or non-spherical particles and even problems in electrodynamics etc. A secondary goal is to demonstrate the effectiveness of utilising volume preserving maps in conjunction with divergence-free interpolation, which we show through numerical tests.
	
 We will now review some standard approaches to these two steps that are available in the literature. The optimal interpolation scheme for use in numerical turbulence has drawn a lot of attention since the 1980s. A number of authors \cite{Deardorff,Riley,Bernard,Portela} use trilinear interpolation methods, however, on marginally resolved grids, there can be significant variations between grid points and Yeung and Pope \cite{yeung_pope_1989} show that trilinear interpolation is not accurate enough for deriving convergent Lagrangian statistics. In the same article Yeung and Pope explore the accuracy of different interpolation methods and conclude that tricubic interpolation is optimal as they best represent the turbulent energy spectrum and provide accurate and smooth approximations to $\bu(\bx,t)$. They also mention, however, that a 13 point Taylor series expansion can be cheaper and accurate enough for practical purposes despite producing discontinuous approximations. Tricubic interpolation is also used by a number of other authors \cite{HOMANN2007}.	In another article, Balachanda and Maxey \cite{BALACHANDA} review a number of methods such as Lagrangian interpolation, partial Hermite interpolation, linear interpolation, a 13 point Taylor series, direct Fourier summation and a shape function method in evaluating fluid velocities from Fourier series. Unsurprisingly, they find that direct Fourier summation is the most accurate but most costly and that linear interpolation is the least accurate but least costly. The remaining methods lie somewhere in between and the authors give recommendations based on the the underlying physical properties of the fluid field. McLaughlin gives a brief review of some current interpolation schemes (see \cite{MCLAUGHLIN1994} and references therein) and suggests that partial Hermite interpolation gives better accuracy over Lagrangian interpolation methods, which is in agreement with \cite{ROVELSTAD1994} who also recommends Hermite methods over others. A disadvantage, however, of partial Hermite interpolation is that one must compute the values of several spatial derivatives in addition to the function itself on an array of points. In practice, this means additional CPU time and memory requirements but can be avoided if accurate values for the derivatives are already available on grid points, for example from some spectral element solvers. \\
	
	The aforementioned methods are now considered amongst the standard procedures for particle tracking and few new schemes have been discussed in such detail since then. Despite the success of these methods, two main drawbacks are present when using, for example, tricubic, Hermite or Lagrangian interpolation polynomials: one is that they are relatively costly compared to other contending methods for example, triquadratic polynomials or 13-point Taylor expansions, which although are less accurate, provide reasonable enough statistics for engineering purposes; another is that they do not respect the divergence-free properties of the fluid field, which has been shown can lead to qualitatively incorrect results \cite{Meyer,Wang,HAM2006}. \\

	Here, we propose the use of RBF interpolation as an approach to address the aforementioned drawbacks. RBFs are a commonly used tool amongst scientist and engineers for approximating data. One of the appealing properties of RBFs for particle tracking is that they are able to provide very accurate vector-valued  approximations that are exactly divergence-free with infinitely many non-vanishing derivatives. In addition, the computation of these approximations require no more computational effort than solving a linear system. A detailed construction and review of RBF interpolation is outside the scope of this paper, a brief introduction will be presented in the following section.\\ 
	
	While a lot of consideration has gone into determining which interpolation scheme to use, the problem of which numerical method to integrate the resulting ODE has not been approached in much detail. All of the above interpolation schemes destroy the divergence-free property of the vector field so it therefore suffices to apply a cheap, all-purpose method as they do in aforementioned references. This is typically achieved through use of a multi-step or multi-stage method such as an Adams method or a Runge-Kutta method. For example, the Adams-Bashforth two-step method is frequently used \cite{mortensen2008dynamics, tapley2018novel, challabotla2015orientation} as this has the advantage of being explicit and only use information of the velocity field at integer multiples of $h$ and hence avoids the need for temporal interpolation of the velocity field.	As the fluid field is inherently divergence-free, it is a logical step to have our pathlines reflect this property. If the underlying vector field is divergence-free, such as those arising from a matrix-valued RBF approximation, then integrating the resulting ODE using a volume preserving will result in a pathlines that preserve volume and can produce qualitatively more accurate results. In this paper we adopt the volume-preserving map of Feng and Shang \cite{Kang1995}. The resulting algorithm, is implicit, however an alternative explicit method is also presented that is not exactly volume-preserving but results in trajectories that are quantitatively similar to the exactly volume-preserving method. The resulting algorithms are able to capture the qualitative features of a helical vortex using far less interpolation points and larger time-steps than a conventional method involving tricubic interpolation and an Adams-Bashforth two-step method.	In the following section we will present the algorithms, the next section presents various numerical experiments and the final section is dedicated to conclusions. 
	
	\section{Numerical methods}
	We begin this section with some considerations when applying a numerical method to an interpolated vector field. We then give a brief introduction to radial basis functions (RBFs) and describe a few of their important features that are relevant to particle tracking. The next section presents the volume preserving method for integrating the ODE and the final section describes the implementation of a benchmark method, which is used as a comparison.
	
	\subsection{Computing pathlines on vector interpolated data}
	This section address the problem of finding accurate solutions to equation \eqref{ODE} by solving 
	\begin{equation}\label{IODE}
	\frac{{\dd \bx}}{\dd t} = \bar{\bu}(\bx,t),
	\end{equation}
	where $\bar{\bu}(\bx,t) = \bu(\bx,t)+\be$ is the interpolated vector field and $\be=\be(\bx,t)$ is the error associated with the interpolation method. Here we see that numerically integrating equation \eqref{IODE} will always see a $\mathcal{O}(||\be||_2)$ discrepancy between the numerical solutions and the true solution of equation \eqref{ODE}. This error term is associated with the interpolation step and cannot be mitigated by increasing the accuracy of the numerical method in a standard way (e.g, increasing the order of the method or reducing the step-size).  In this sense, the interpolation accuracy places a bound on the total accuracy of the resulting algorithm. It is therefore unwise to use too small an integration step-size as this will result in convergence to the wrong trajectory. On the other hand, using an extremely accurate interpolation scheme and integrating the resulting vector field using an inaccurate method is a waste of computational resources, as the effort gone into minimising $\be$ will be polluted by the global error of the ODE solver. In this sense, one should choose the interpolation method with the ODE solver in mind to design the an efficient algorithm. It makes sense to choose a step-size such that the error from the ODE step is of the same magnitude as the interpolation error. That is, the choice of $h$ should roughly satisfy 
	 \begin{equation}
	 	\bdelta(h)\ge \be,
	 \end{equation} 
	 where $\bdelta(h)$ is the $h$-dependent local error of the numerical method. Decreasing $h$ below this bound will not resolve in more accurate solutions. The value of $h$ where $\bdelta(h) = \be$ will be henceforth referred to as the ``saturation point" and values of $h$ where $\bdelta(h) \le \be$ the ``saturation region" in which the local error is dominated by the interpolation error. \\
	
	To illustrate this concept using a concise example, consider integrating equation \eqref{IODE} with the forward Euler method, given by
	\begin{equation}
	{\bx}_{i+1} =   {\bx}_i + h \bar{\bu}({\bx}_i,t) \label{ife},
	\end{equation}
	where $\bx_{i+1}$ is the numerical approximation to the exact solution of equation \eqref{ODE} $\bx(t_i+h)$. The local error $\bdelta_{i+1} = \bx_{i+1} - \bx(t_i+h)$ is then computed by Taylor expanding the exact solution $\bx(t_i+h)$ about $t_i$. Inserting $\bx(t_i)=\bx_i$, we arrive at
		\begin{align}
	{\bdelta}_{i+1} = &  h ( \bar{\bu}(\bar{\bx}_i,t)-\bu(\bx_i,t)) + \frac{ h^2}{2}\nabla \bu(\bx_i,t) \, \bu(\bx_i,t) + \mathcal{O}(h^3),\nonumber\\
	= & h\be + \frac{ h^2}{2}\nabla \bu(\bx_i,t) \, \bu(\bx_i,t) + \mathcal{O}(h^3),
	\end{align}
	which implies that 
	\begin{equation}
		 ||{\bdelta}_{i+1}||_2 \le  h ||\be||_2+C h^2,
	\end{equation}
	 where the constant $C$ only depends on the derivatives of the true vector field $\bu(\bx,t)$. It then follows that after a sufficient number of time-steps, the global error turns out to be of order $\mathcal{O}(||\be||_2)+\mathcal{O}(h)$. Clearly we reach a point where reducing $h$ will not decrease the error as the $\mathcal{O}(||\be||_2)$ term will dominate. In general, expressions for $\be$ are not known for most interpolation methods, however approximations and bounds are available in the literature. For example, order $q-1$ polynomial interpolation has error $\mathcal{O}(\Delta x^q)$ for grid spacing $\Delta x$, so for an order $p-1$ numerical method with local error of order $\mathcal{O}(h^p)$, the saturation region is characterised by
	\begin{equation}
		h\le D \Delta x^{\frac{q}{p}},
	\end{equation}
	for some constant $D$ that may depend on $\bu(\bx,t)$ and its derivatives but not $h$ or $\Delta x$.
	
	\subsection{Step 1: Spatial interpolation using matrix radial basis functions}
	In this section we are faced with the problem of creating an continuous divergence-free approximation to a set of discrete data points $\{\bx_i,d_i\}_{i=1}^{N}$, where $d_i=d_i(t_0)$ is the value of the data at some time $t_0$ and at the location $\bx_i$, which corresponds to a grid node. RBF interpolation differers from classical polynomial interpolation in that the interpolating surface is a linear combination of a positive definite radial functions $\psi(r_i)$ that is centred at a grid node $\bx_i$, and depends only on the distance $r_i=||\bx-\bx_i||_2$ from that node. Such a surface is represented by
	\begin{equation}
	s(\bx) = \sum_{i=1}^{N} \psi(|| \bx - \bx_i ||_2) c_i,
	\end{equation}
	where the constants $c_i$ are chosen such that surface is consistent with the data points $s(\bx_i) = d_i$. This is done by solving the linear system 
	\begin{equation}
	A \mathbf{c} = \mathbf{d},
	\end{equation}
	where $A_{ij} = \psi(|| \bx_i - \bx_j ||_2)\in\mathbb{R}^{N\times N}$ is a positive definite matrix. In this way we can construct a vector valued interpolating surface by interpolating each component of the fluid field independently. Fluid fields inherently satisfy the condition $\nabla\cdot\bu(\bx,t)=0$ so it would make sense that our interpolating surface also satisfies this quality, however the surface constructed from the above scalar RBFs formalism is not guaranteed to satisfy $\nabla\cdot\bar{\bu}(\bx,t_0)=0$. This is easily remedied through use of \textit{matrix valued} RBFs. In a similar fashion to the above scalar RBFs, constructing a matrix RBF interpolant involves solving a linear system to find a set of, now vector valued, coefficients. In this way, we can calculate the three components of the interpolating surface $\bar{\bu}(\bx,t_0)$ simultaneously and have that they define a divergence-free field. First, we define a matrix-valued radial basis function by 
	\begin{equation}
	\Phi(r) = (\nabla^T \nabla - \Delta \mathbb{1})\psi(r)
	\end{equation}
	for some scalar radial basis function $\psi(r)$. Then the vector valued interpolating surface is constructed by 
	\begin{equation}
	\bar{\bu}(\bx,t_0) = \sum_{i=1}^{N} \Phi(|| \bx - \bx_i ||) \bc_i.
	\end{equation}
	 Taking the divergence of $\Phi(r) \bc_i$ and with the aid of some vector identities we arrive at
	 \begin{align}
	 	\nabla\cdot\Phi(r) \bc_i = & \nabla\cdot(\nabla^T \nabla - \Delta \mathbb{1})(\psi(r) \bc_i), \nonumber\\
	 	=& \nabla\cdot(\nabla \times (\nabla \times(\psi(r) \bc_i)),\nonumber\\
	 	=&0
	 \end{align}
	 as the divergence of curl is zero. It then follows that $\nabla\cdot\bar{\bu}(\bx,t_0)=0$. The vector coefficients $\bc_i$ are chosen such that $\bar{\bu}(\bx_i,t_0) = \bu(\bx_i,t_0)$ which amounts to solving a single $3 N \times 3 N$ linear system for the $N$ vector values coefficients $\bc_i$ (as opposed to the scalar case where we solve three $N \times N$ linear systems). For more details on RBF interpolation we refer to  \cite{buhmann2003radial,lowitzsch2004approximation,McNally}. \\
	
	Henceforth, we will use a radial basis function known as inverse quadrics, given by 
	\begin{equation}
		\psi(r) = \frac{1}{1+(\epsilon r)^2},
	\end{equation}
	where $\epsilon$ is called the shape parameter and determines the ``flatness" of $\psi(r)$. In general, one should choose $\epsilon$ as low as possible, which results in more accurate representations of the data. \\
	
	In addition to a more accurate interpolating surface, RBF interpolation has the advantage of approximating the vector field, with a $C^\infty$ surface that has infinitely many non-vanishing derivatives. This means that we can find good approximations of derivatives by simply evaluating the derivative of the interpolating surface, which is required in more complicated particle models, for example tracking non-spherical inertial particles \cite{mortensen2008dynamics,tapley2018novel,challabotla2015orientation}. In addition, we are not restricted to using a particular number of interpolation points. In this way we can match the accuracy of the interpolation step to the accuracy requirements of the ODE solver and hence, is more accommodating when optimising the choice of $h$ and $\Delta x$. We are not afforded this freedom with a typical polynomial method, which often requires solving a linear system of fixed size to ensure the existence of a unique interpolating polynomial.

	\subsection{Step 2: Integration using a volume preserving map}
	As the resulting vector field is divergence-free it is a logical step to preserve this feature by applying a volume preserving method. While it has been shown that generic B-series methods cannot be exactly volume preserving \cite{Iserles2007} there exist some Runge-Kutta methods that instead preserve a modified measure \cite{BADER2016}, as well as exactly volume-preserving methods that were discovered by Quispel \cite{QUISPEL1995} and Feng and Shang \cite{Kang1995}. It is not clear as to whether one method is better than the other, as they are both implicit and involve the solving an integral. Here, we will adopt the method of Feng and Shang and refer to \cite{Kang1995,GNI} for a detailed analysis and construction of the method. The method begins by splitting the ODE into two sub-systems 
	\begin{equation}
	\dot{\bx,t} = \bar{\bu}(\bx,t) = \bu_{1}(\bx,t)+\bu_{2}(\bx,t),
	\end{equation}
	where $\bar{\bu}(\bx,t)=(u,v,w)^\mathrm{T}$ is now a matrix RBF vector field that is a known function of space at a particular time and $\bu_{1}(\bx,t)$ and $\bu_{2}(\bx,t)$ are the Hamiltonian vector fields
	\begin{equation}
	\bu_{1}(\bx,t) = \left( u , -\int_{0}^{y}u_x \dd y , 0 \right)^\mathrm{T} 
	\quad\mathrm{and}\quad
	\bu_{2}(\bx,t) = \left(0, v + \int_{0}^{y}u_x \dd y, w \right)^\mathrm{T} 
	\end{equation}
	whose flows respectively preserve the Hamiltonians
	\begin{equation}
	H_{1} = \int_{0}^{y}u \,\dd y \quad\mathrm{and}\quad H_{2} = \int_{0}^{z}\left( \frac{\partial H_1}{\partial x} - v \, \right)\dd z + \int_{0}^{y} w|_{z=0} \dd y.
	\end{equation} Note that we now have to evaluate an integral. In our case this amounts to taking the integral of a linear combinations of shifted inverse quadrics functions, which can be done exactly and results in non-separable and rational polynomial Hamiltonians. We now have the two Hamiltonian ODEs
	\begin{equation}
	\dot{\bx}_1 = \bu_{1}(\bx_1)
	\quad\mathrm{and}\quad
	\dot{\bx}_2 = \bu_{2}(\bx_2).
	\end{equation}
	Feng and Shang show that a splitting method based on the above vector fields is volume preserving if the numerical flows of the split vector fields, denoted by $\bvphi_{h}^{[1]}(\bx_1)$ and $\bvphi_{h}^{[2]}(\bx_2)$, preserve symplecticity. We will use the implicit midpoint rule, which is known to be a symplectic map. The numerical flow of the original ODE is now computed by the second-order Strang splitting operator
	\begin{equation}\label{splitting}
	\bphi_h = \bvphi_{h/2}^{[1]}\circ\bvphi_{h}^{[2]}\circ\bvphi_{h/2}^{[1]},
	\end{equation}
	which can be thought of as a composition of sub-flows of area-preserving maps. The main result of \cite{Kang1995} is that 
	\begin{equation}
	\left|\frac{\partial \bphi_h(\bx)}{\partial \bx}\right| = 1
	\end{equation}
	and hence $ \bphi_h(\bx^n)$ preserves volume. 
	
	\subsubsection{Explicit method}
	Whilst volume-preservation is a desirable solution quality, the resulting algorithm is implicit, which is a costly feature of the method. It is not clear that a method needs to be \textit{exactly} volume preserving for application purposes and in many cases, cost-effectiveness is a more favourable solution feature than qualitative accuracy.  An alternative to above method is to replace the implicit mid-point step with an \textit{explicit} mid-point step, which is given by 
	\begin{equation}
	\bvphi_{h}^{[i]}(\bx_i^n) = \bx^{n} + h \bu_i(\bx_i^n+\frac{h}{2}\bu_i(\bx^n_i)).  
	\end{equation}
	The explicit mid-point method is not symplectic and therefore the algorithm will not preserve volume, so this method can be thought of as an explicit approximation to a volume-preserving method and we will show through numerical tests that this method performs surprising well.	
	
	\subsection{Benchmark method: tricubic interpolation}
	 As mentioned in the introduction, there are a few contending interpolation methods that are frequently used in application depending on the availability of fluid derivatives and the accuracy and cost requirements. We will focus on the most general setting where only the fluid values are available on the grid points and therefore consider tricubic interpolation as a reasonable  benchmark interpolation method given its superior accuracy over 13 point Taylor expansions, Lagrange polynomials and linear schemes, for example. Tricubic polynomial interpolation also has the advantage of being fourth-order accurate in space and $\mathcal{C}^1$ continuous across grid cell boundaries.	Here we will give a brief description, but for a detailed report on the construction, implementation and analyis of the following method, we refer the reader to \cite{Marsden2005}. Tricubic interpolation polynomials are of the form
	\begin{equation}
	s(\bx) = \sum_{i,j,k=0}^{3}a_{ijk} x^iy^jz^k,
	\end{equation}
	where the problem is to find the $64$ $a_{ijk}$ coefficients such that $s(\bx_i)=d_i$, given a set of data points $\{\bx_i,d_i\}_{i=1}^{64}$, where each $\bx_i$ is a vertex of the $4\times 4 \times 4$ grid that is centered at $\bx$. By stacking the $a_{ijk}$ coefficients into the vector $\ba \in \mathbb{R}^{64}$, we can form the following linear system 
	\begin{equation}
	M \ba = \bd,
	\end{equation}
	where each row of $[M_{mn}]$ are the values of the monomials $x^iy^jz^k$ corresponding to the $a_n$ coefficient and evaluated at the location of $d_m$. It can be shown that method is equivalent to performing many 1D cubic interpolations in different dimensions and evaluating it at $\bx$. The resulting field is $\mathcal{C}^\infty$ smooth locally, but across a cell face we cannot have more than $\mathcal{C}^1$ continuity. To construct a vector valued interpolating surface, each vector component is calculated separately and hence requires the solution to three $64\times 64$ linear systems. The resulting vector field is then integrated using the explicit Adams-Bashforth two-step method. 
	
	\section{Numerical experiments}
	In this section, we are calculating pathlines of the following helical Taylor-Green flow field
	\begin{align}
	u(x,y,z) =& \sin(x)\cos(y)f(t),\\ 
	v(x,y,z) =& -\cos(x)\sin(y)f(t),\\ 
	w(x,y,z) =& 1 ,
	\end{align} 
	 from the initial position $\bx(0) = (\frac{1}{\sqrt{2}},\frac{1}{\sqrt{2}},\frac{1}{10})^\mathrm{T}$. Here $f(t) = (1+\dfrac{\sin(\pi\,t/50)}{2})$ is used to give the field time-dependence. We assert that a good qualitative measure of the solution accuracy is the ability for the method to reproduce a helix with bounded radius. We compare four solution methods which is summarised in table \ref{table}. In addition to the MRBF+VP and the MRBF+EMP methods, we will also examine the performance of an MRBF+AB method to isolate the advantages of RBF interpolation over tricubic interpolation.  \\

	\begin{table}
			\hspace*{-2cm}
			\begin{tabular}{c||c|c||c|c}
				Abbreviation & Interpolation & \begin{tabular}{c}
					Div- \\ 
					free
				\end{tabular} & Integration & \begin{tabular}{c}
					Volume \\ 
					preservation
				\end{tabular} \\
				\hline
				MRBF+VP & Matrix RBFs & yes & Strang splitting + implicit mid-point& yes\\		
				MRBF+EMP & Matrix RBFs& yes & Strang splitting + explicit mid-point& no\\
				MRBF+AB2 & Matrix RBFs& yes & Adams-Bashforth two-step& no\\
				TC+AB2 & Tricubic & no & Adams-Bashforth two-step& no\\
				Reference & Exact evaluation & - & Fourth order Runge-Kutta& - \\
			\end{tabular}

		\caption{Summary of the four numerical methods and the reference solution.}\label{table}
	\end{table}
	\begin{figure}
		\centering
		\includegraphics[width=0.5\linewidth]{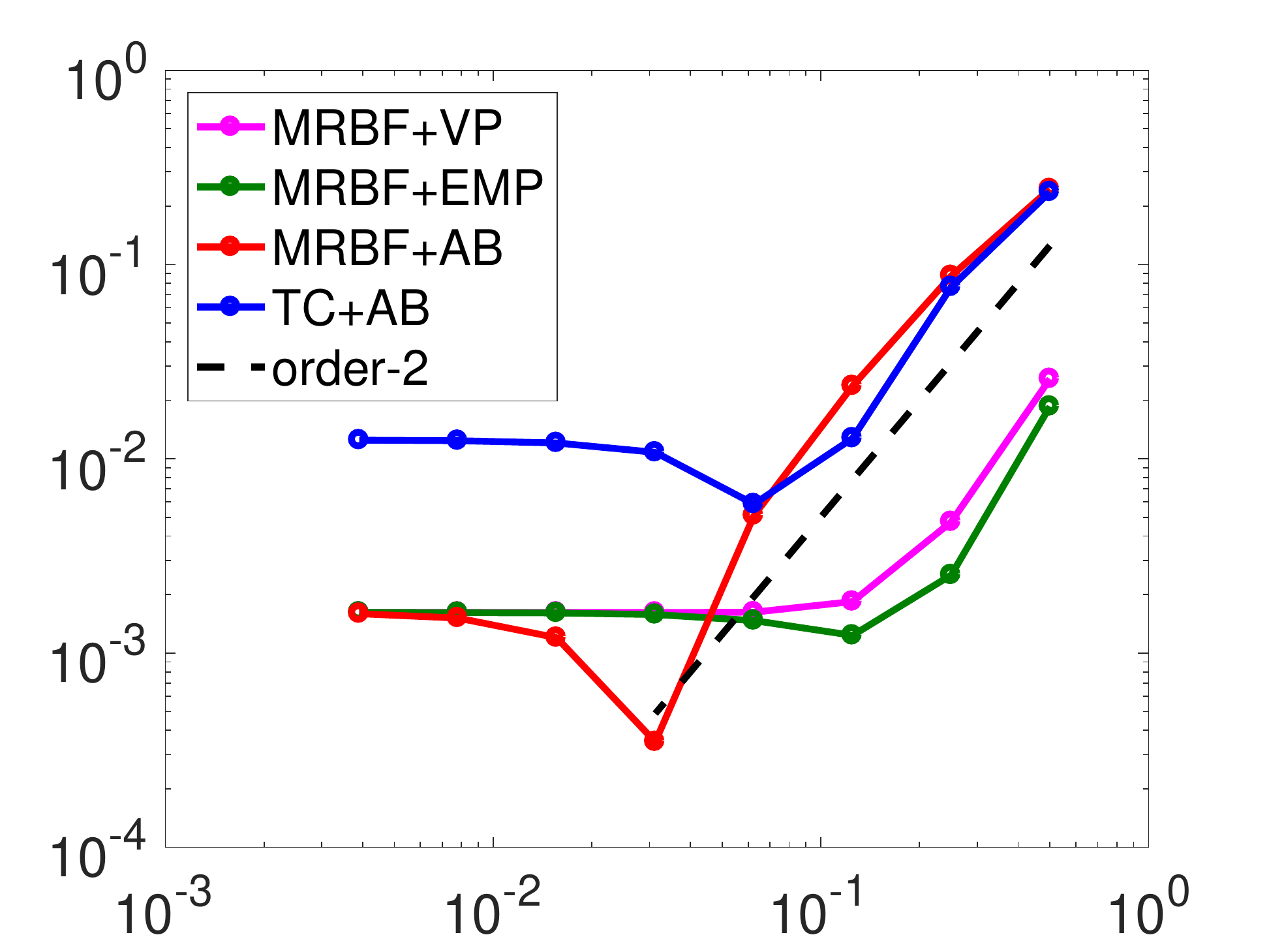}
		\caption{The 2-norm relative global errors at $T=10$ vs step-size $h$.}\label{globalerr}
	\end{figure}
	Figure \ref{globalerr} shows the relative global error convergence of the four methods at $t=10$. The error is calculated from 
	\begin{equation}
		\frac{||\bx_n-\bx^{\mathrm{ref}}_n||_2}{||\bx^{\mathrm{ref}}_n||_2},
	\end{equation}
	 where $\bx_n^\mathrm{ref}$ is the reference solution and $\bx_n$ is the numerical solution. Here the matrix RBF interpolation uses the $4\times4\times4$ nearest grid points that are located on a regular grid with a spacing of $\Delta x = 1/2$ in each direction. We observe roughly order-two convergence for high step-sizes and that the total error is polluted by the interpolation error as the step-size approaches the saturation point. At lower step sizes, we enter the saturation region where the error is dominated by the interpolation error as seen by the, $h-$independent line. Note that we get some cancellation between the interpolation error and the integration error at the saturation point which is seen as a dip in error below the interpolation error. Another observation to be made here is that the interpolation error in the saturation region is much lower for the three matrix RBF solutions than the tri-cubic solution. In addition, in the region where order-two convergence is achieved, the MRBF+VP and the MRBF+EMP errors are about an order more accurate than the Adams-Bashforth solution.  \\
	
\begin{figure}
		\centering
		\begin{subfigure}[b]{0.25\textwidth}
			\includegraphics[width=\linewidth]{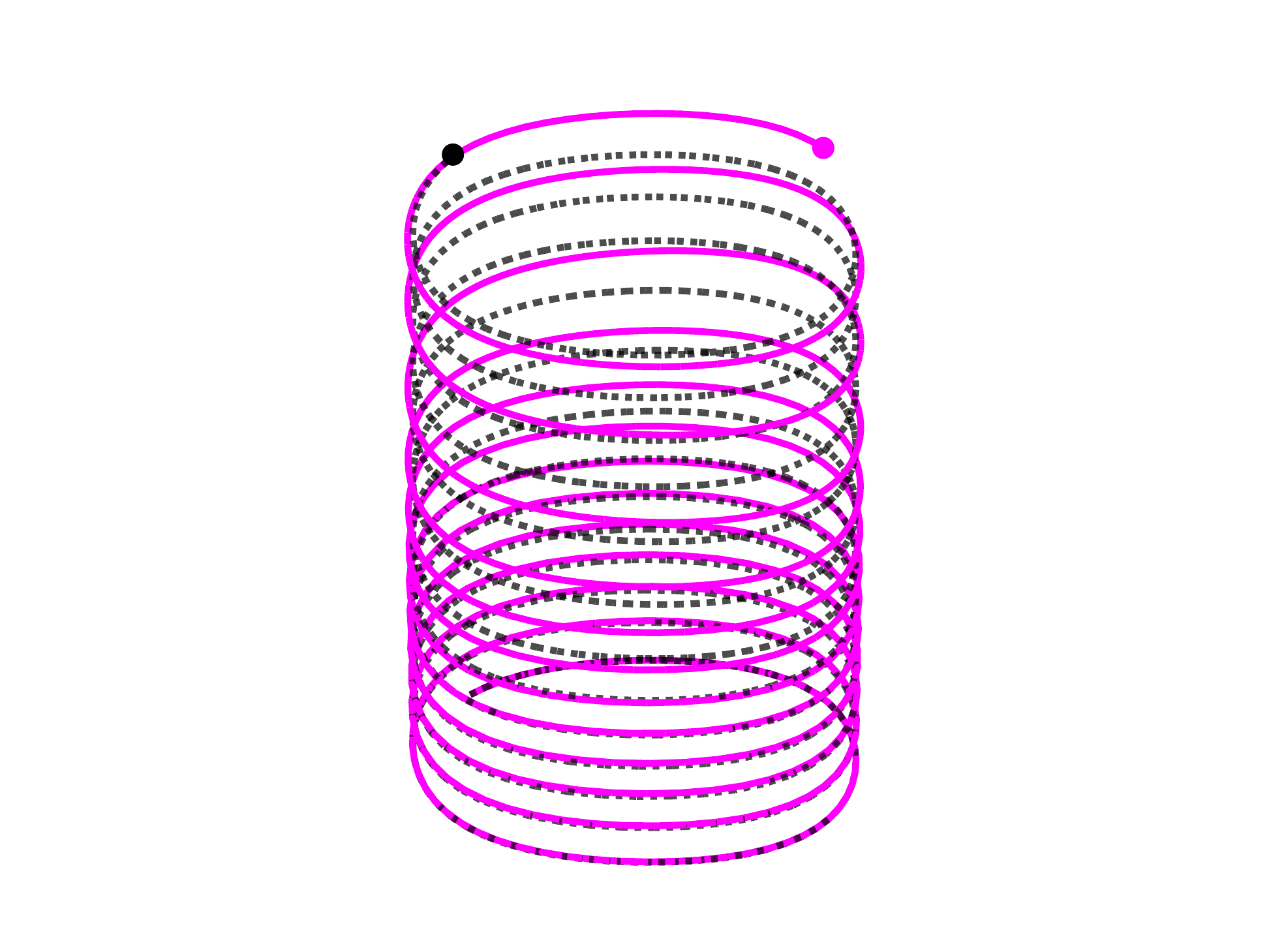}
			\caption{}
		\end{subfigure}
		\hspace{-1cm}
		\begin{subfigure}[b]{0.25\textwidth}
			\includegraphics[width=\linewidth]{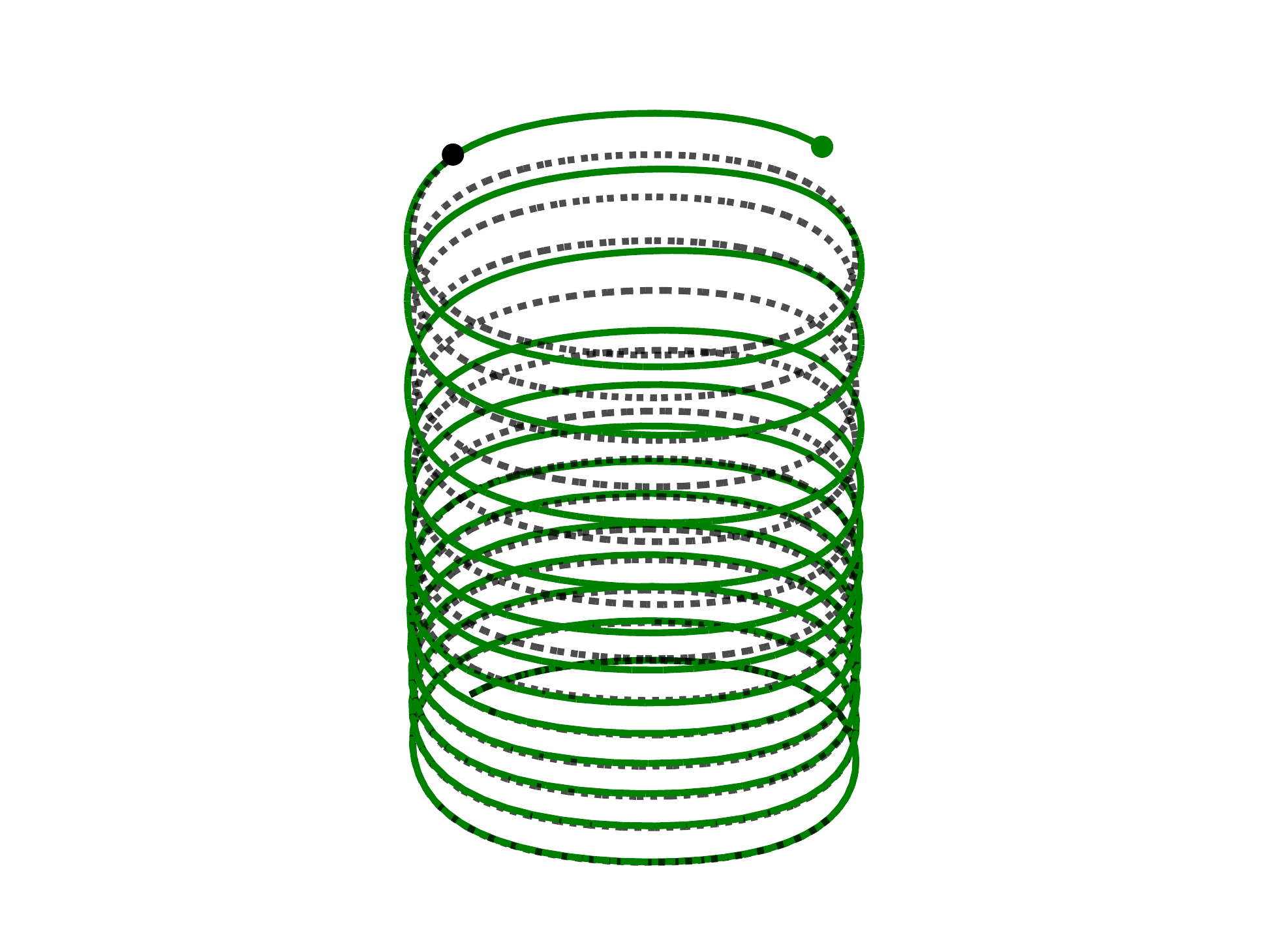}
			\caption{}
		\end{subfigure}
		\hspace{-1cm}
		\begin{subfigure}[b]{0.25\textwidth}
			\includegraphics[width=\linewidth]{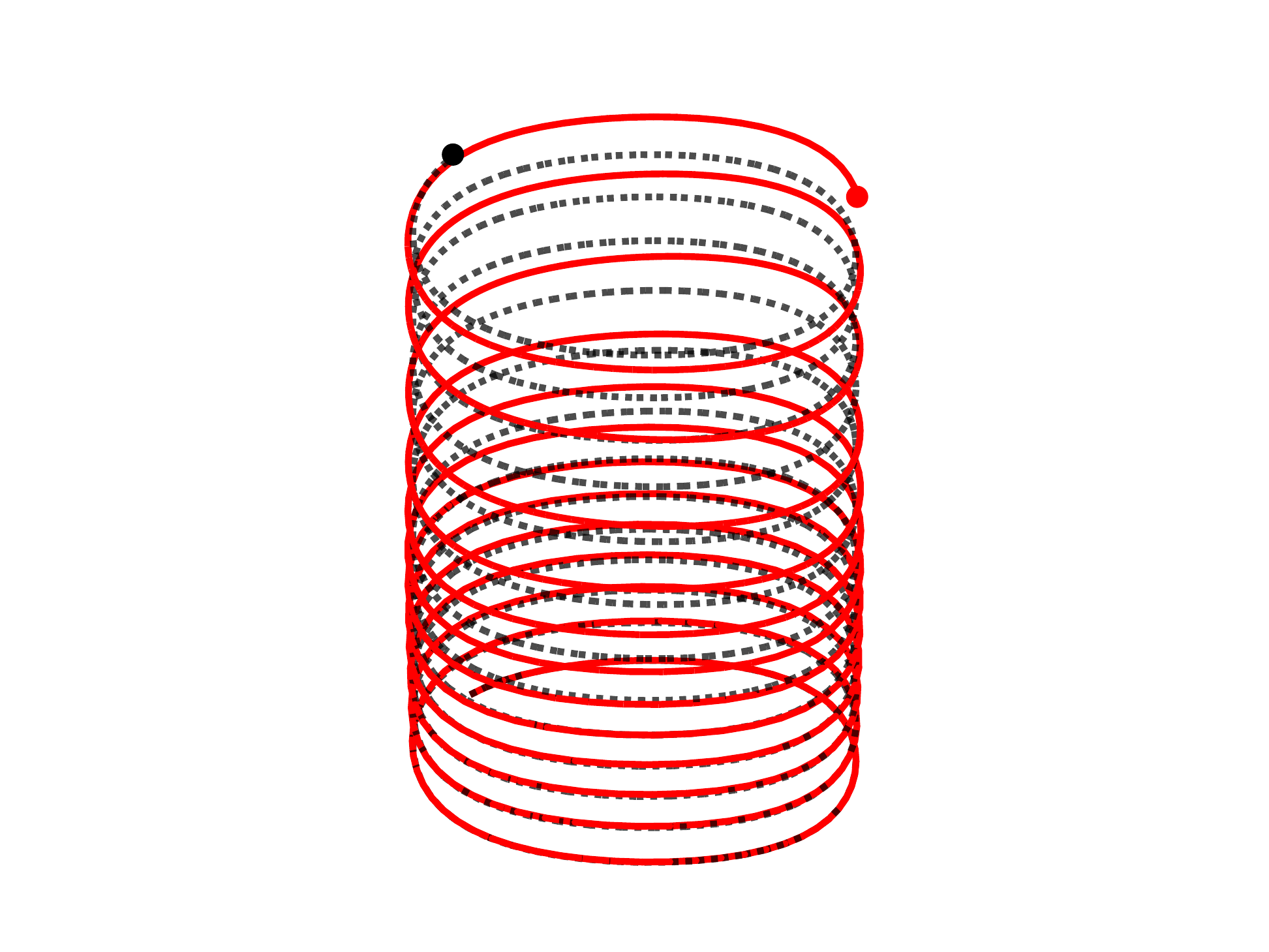}
			\caption{}
		\end{subfigure}
		\hspace{-1cm}
		\begin{subfigure}[b]{0.25\textwidth}
			\includegraphics[width=\linewidth]{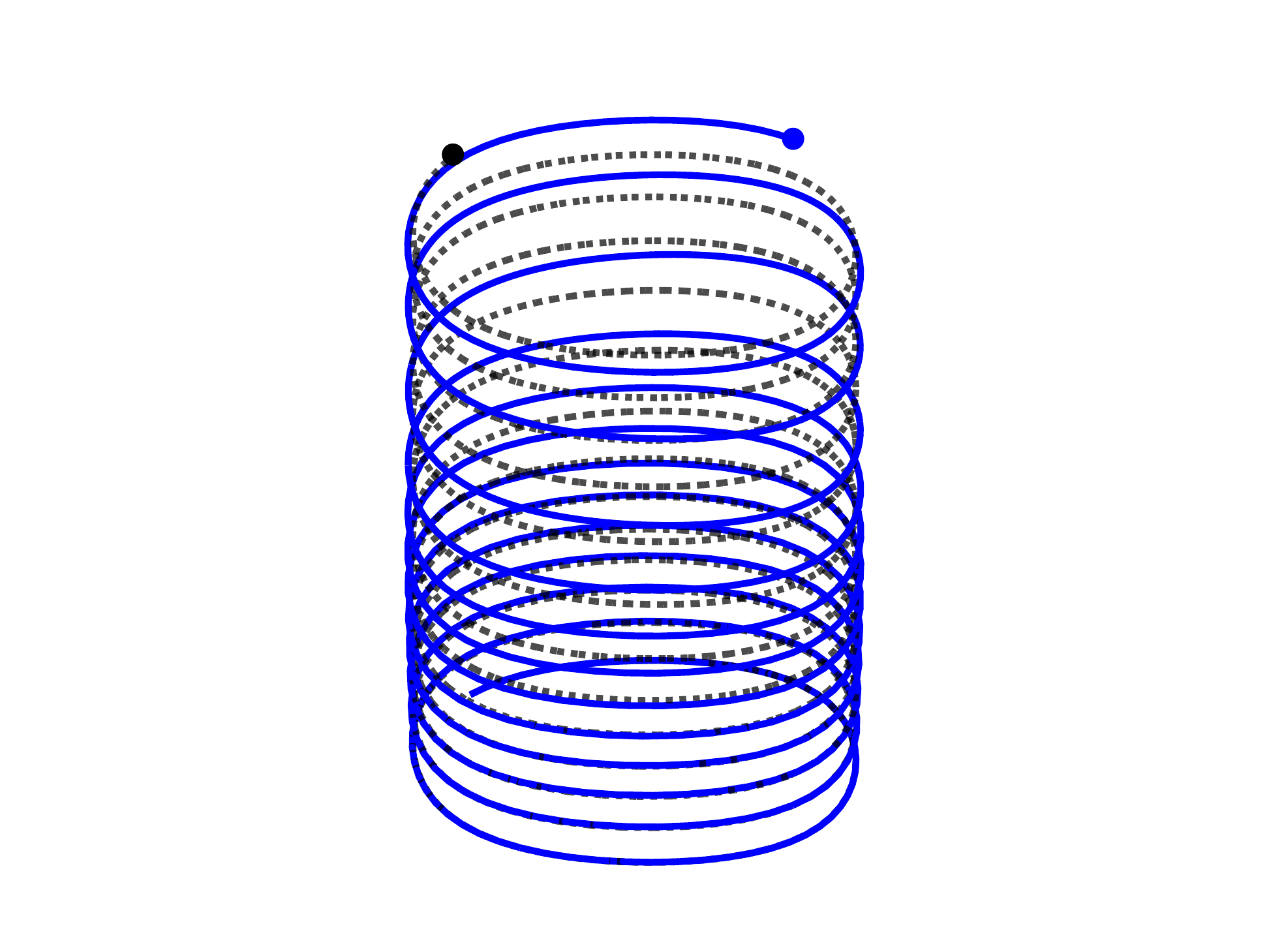}
			\caption{}
		\end{subfigure}\textbf{}
		\begin{subfigure}[b]{0.25\textwidth}
			\includegraphics[width=\linewidth]{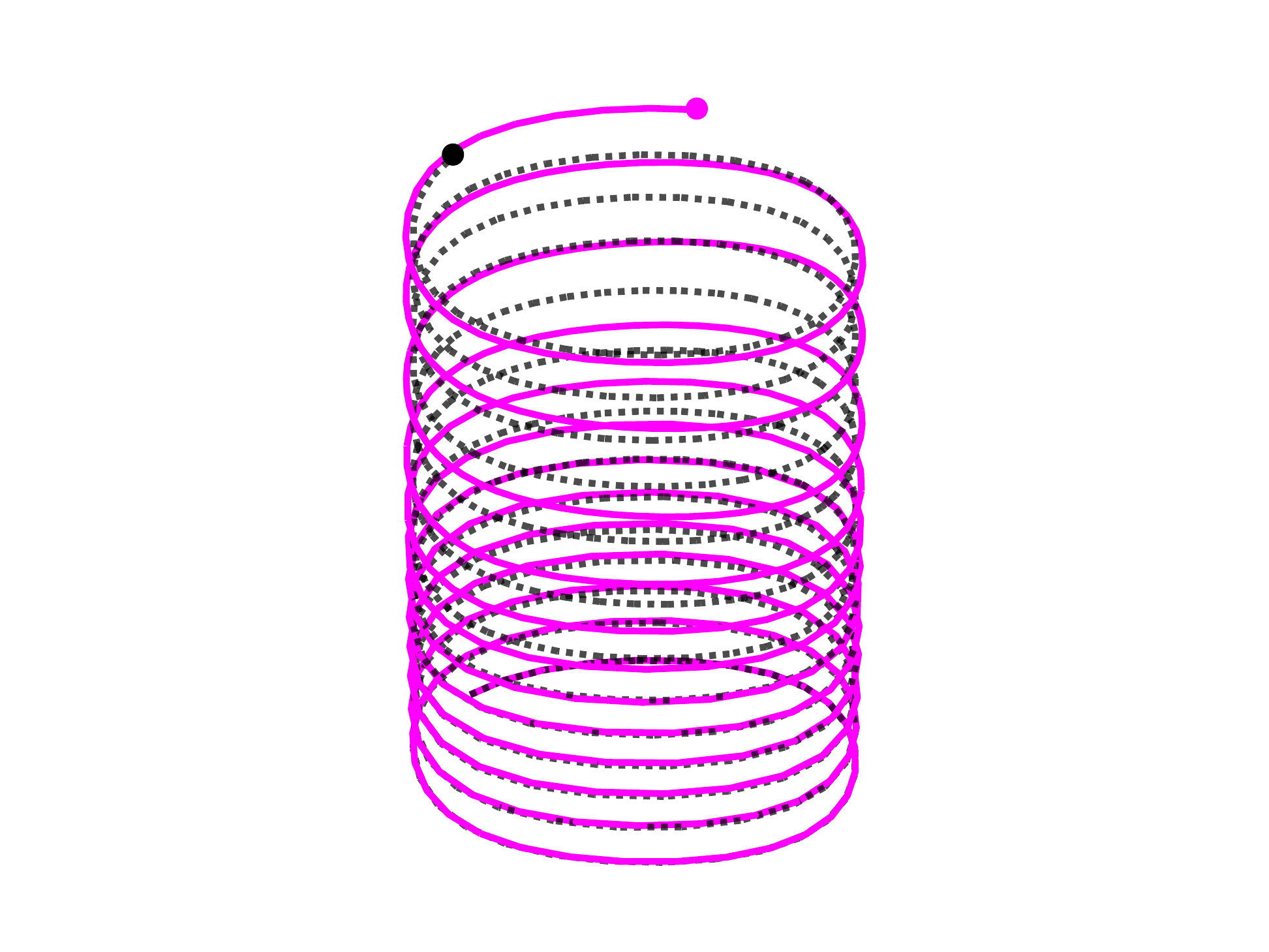}
			\caption{}
		\end{subfigure}
		\hspace{-1cm}
		\begin{subfigure}[b]{0.25\textwidth}
			\includegraphics[width=\linewidth]{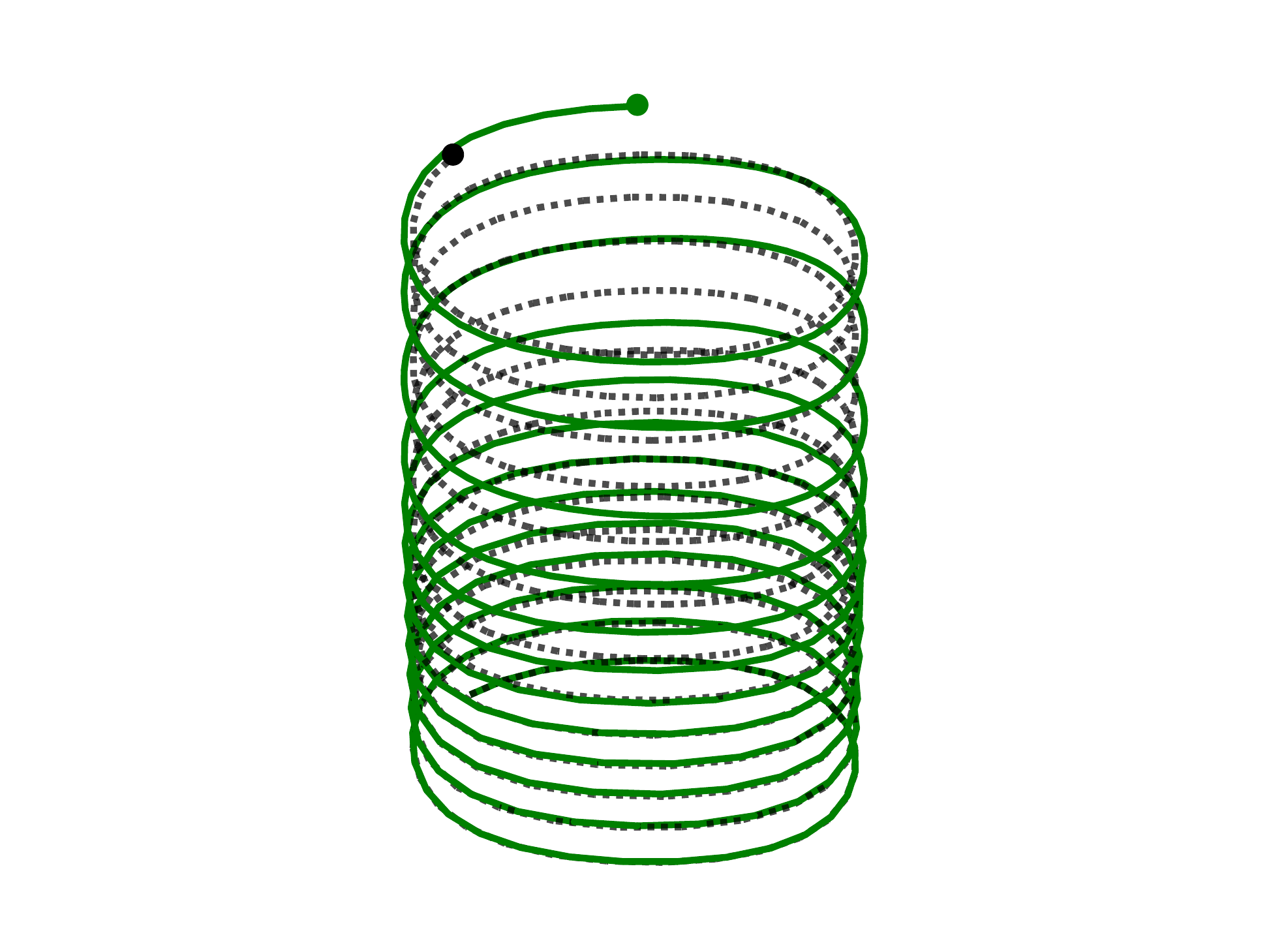}
			\caption{}
		\end{subfigure}
		\hspace{-1cm}
		\begin{subfigure}[b]{0.25\textwidth}
			\includegraphics[width=\linewidth]{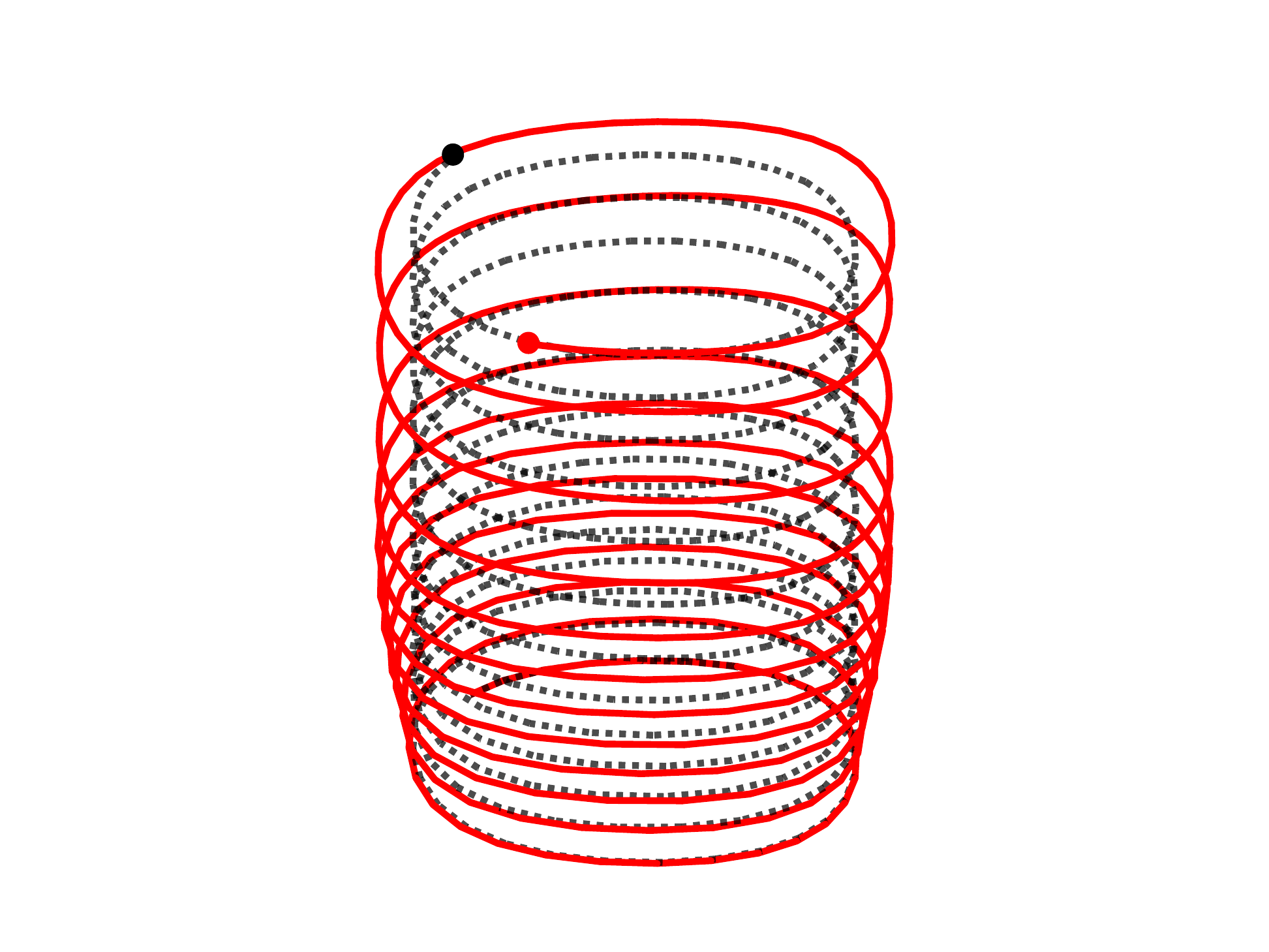}
			\caption{}
		\end{subfigure}
		\hspace{-1cm}
		\begin{subfigure}[b]{0.25\textwidth}
			\includegraphics[width=\linewidth]{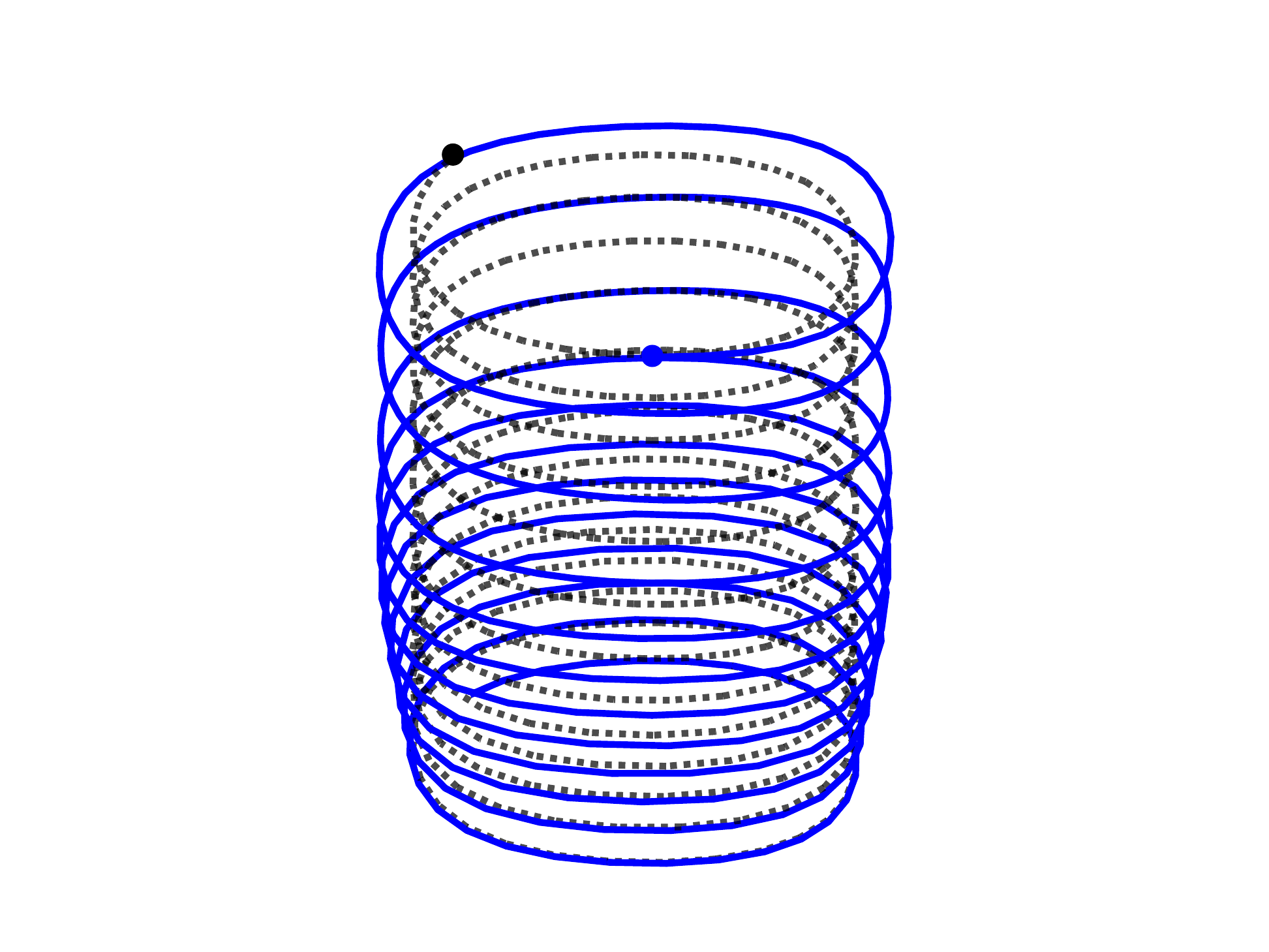}
			\caption{}
		\end{subfigure}
		\begin{subfigure}[b]{0.25\textwidth}
			\includegraphics[width=\linewidth]{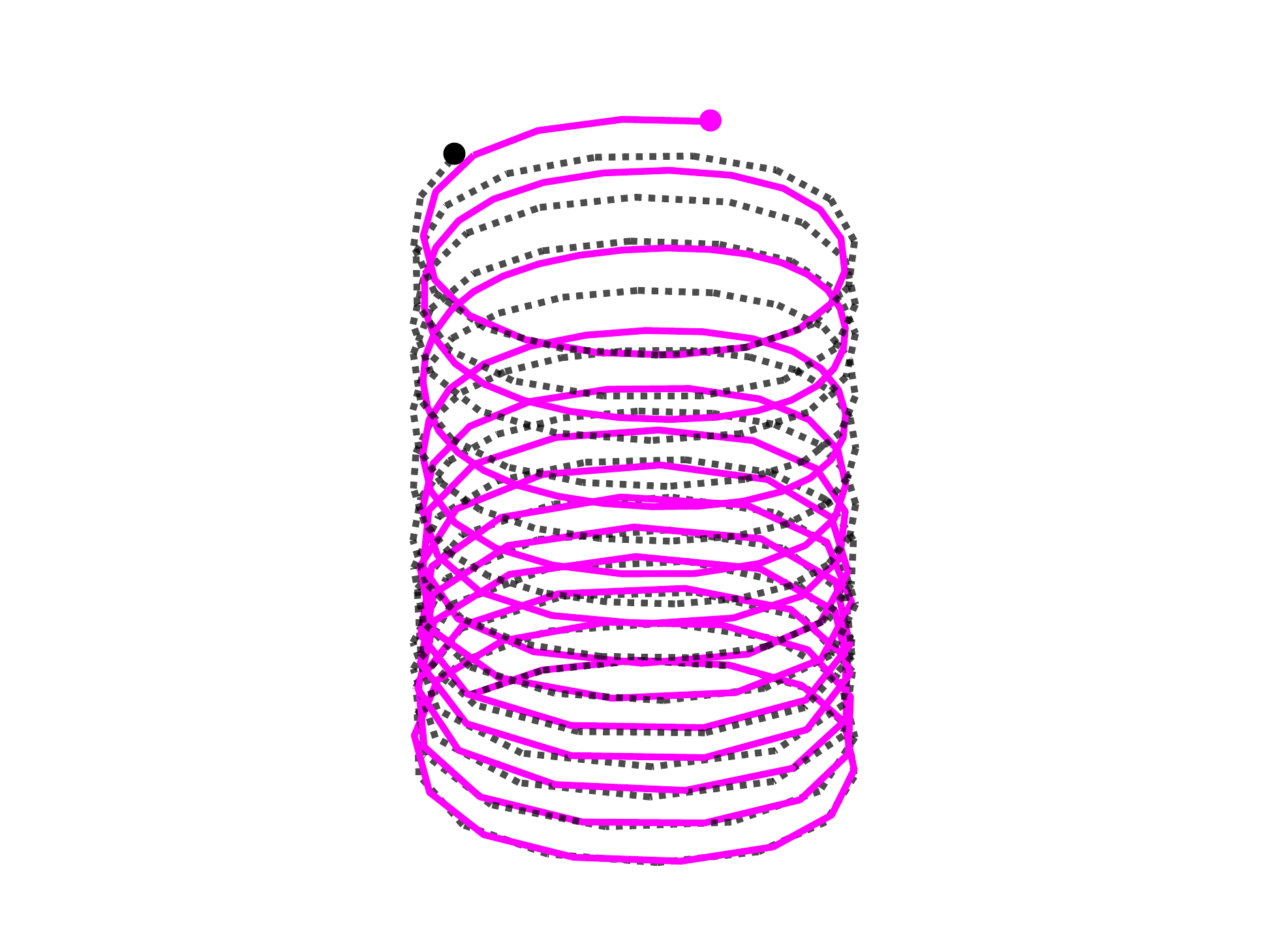}
			\caption{}
		\end{subfigure}
		\hspace{-1cm}
		\begin{subfigure}[b]{0.25\textwidth}
			\includegraphics[width=\linewidth]{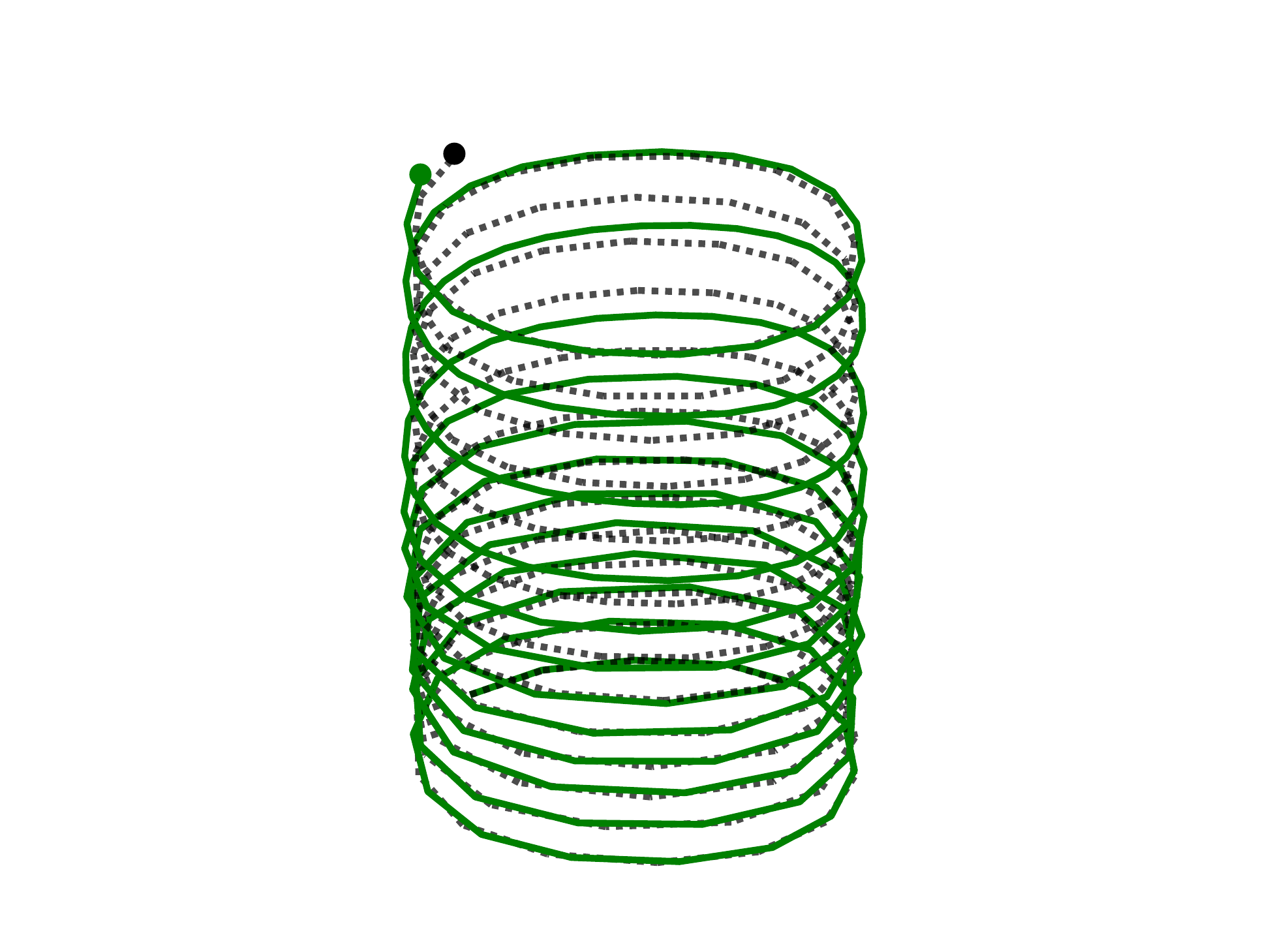}
			\caption{}
		\end{subfigure}
		\hspace{-1cm}
		\begin{subfigure}[b]{0.25\textwidth}
			\includegraphics[width=\linewidth]{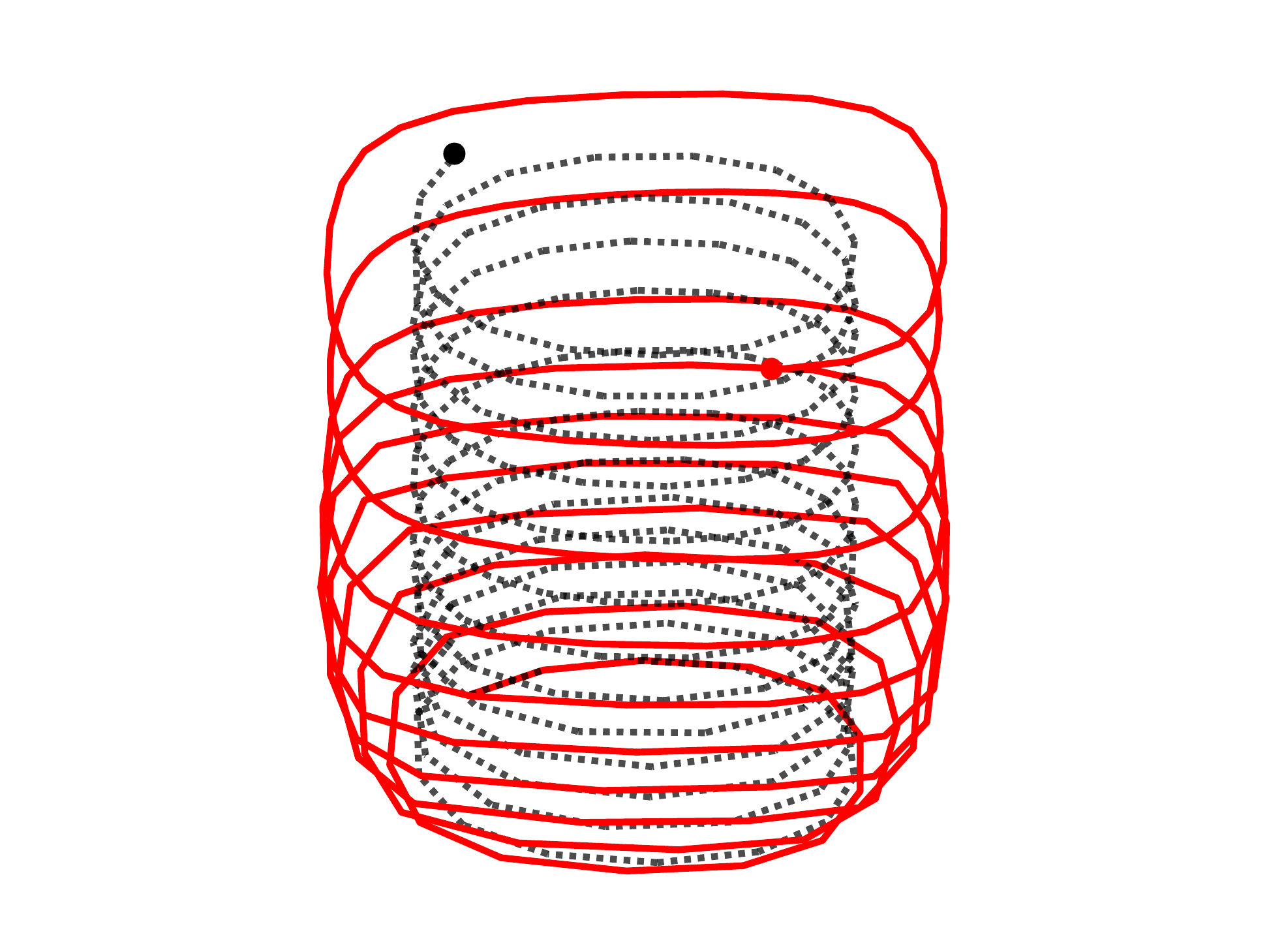}
			\caption{}
		\end{subfigure}
		\hspace{-1cm}
		\begin{subfigure}[b]{0.25\textwidth}
			\includegraphics[width=\linewidth]{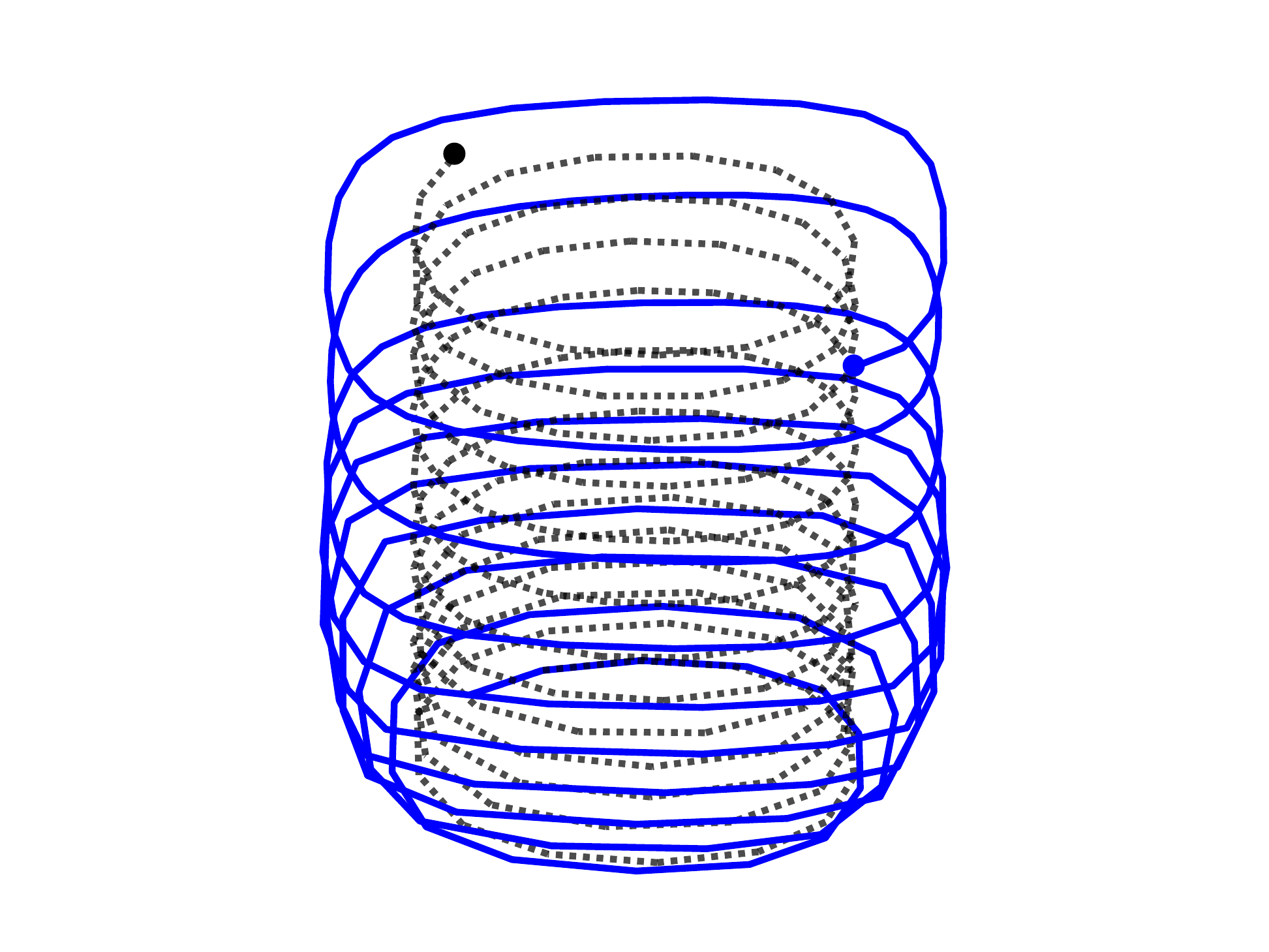}
			\caption{}
		\end{subfigure}
		\begin{subfigure}[b]{0.25\textwidth}
			\includegraphics[width=\linewidth]{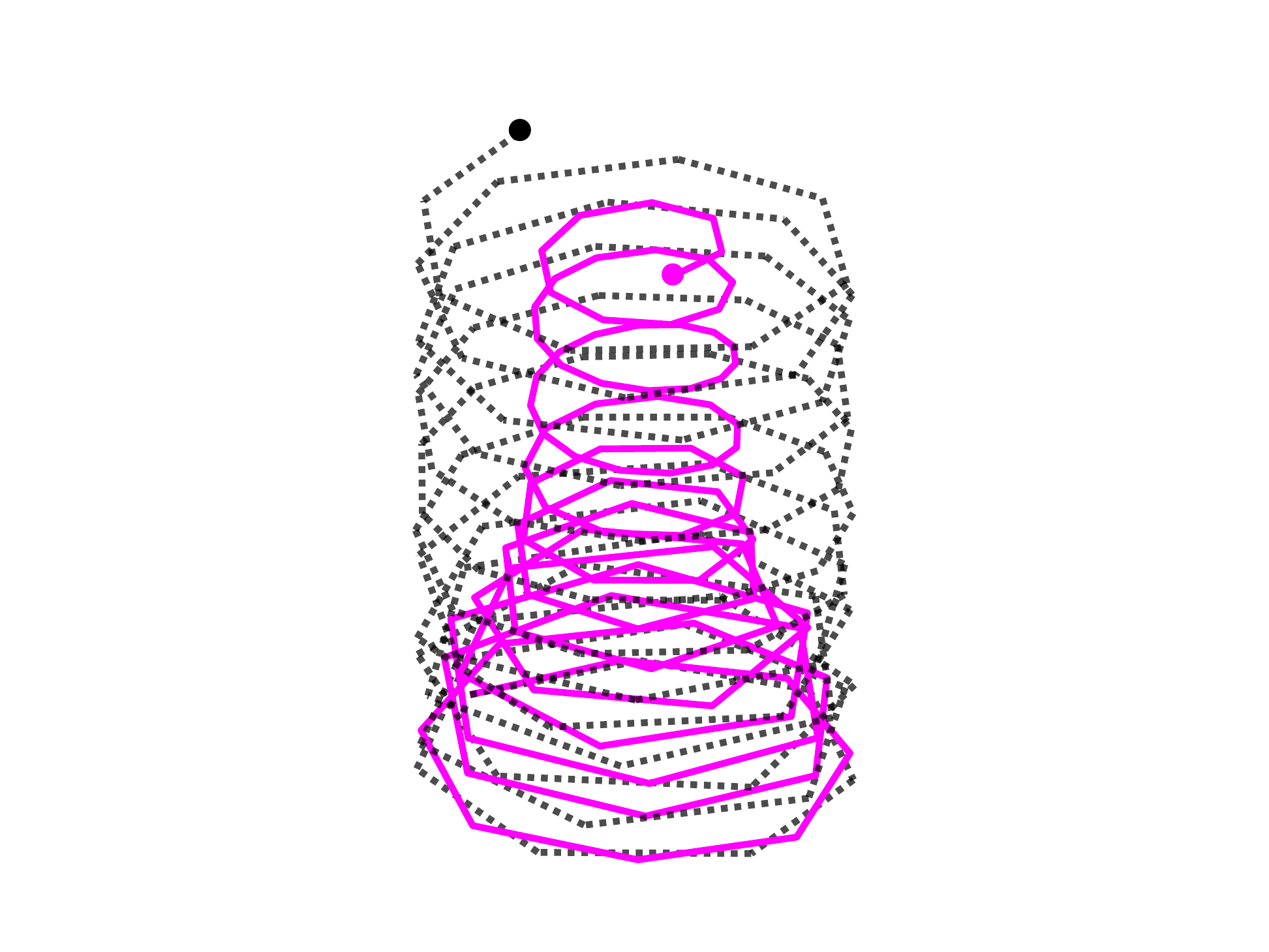}
			\caption{}
		\end{subfigure}
		\hspace{-1cm}
		\begin{subfigure}[b]{0.25\textwidth}
			\includegraphics[width=\linewidth]{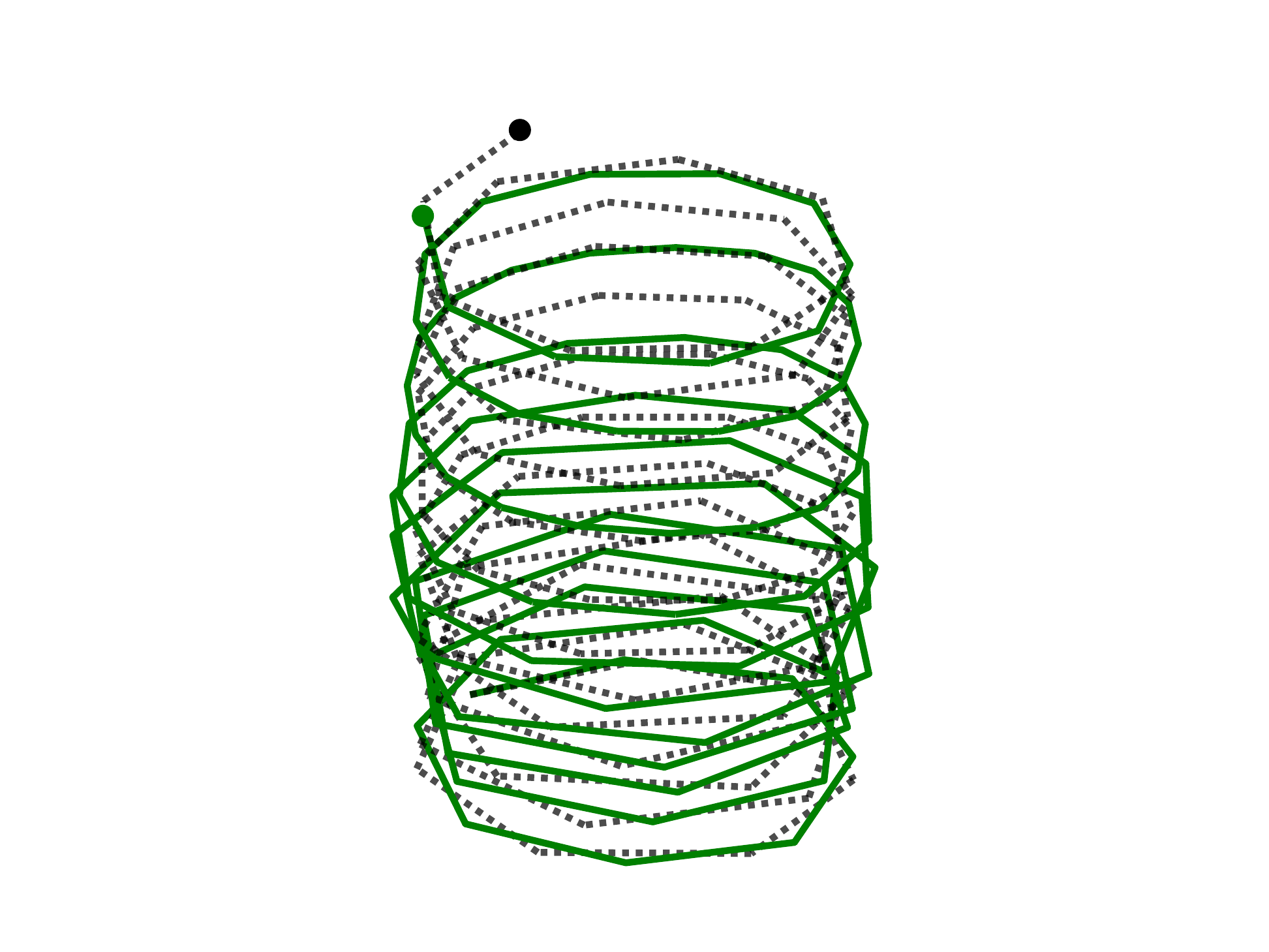}
			\caption{}
		\end{subfigure}
		\hspace{-1cm}
		\begin{subfigure}[b]{0.25\textwidth}
			\includegraphics[width=\linewidth]{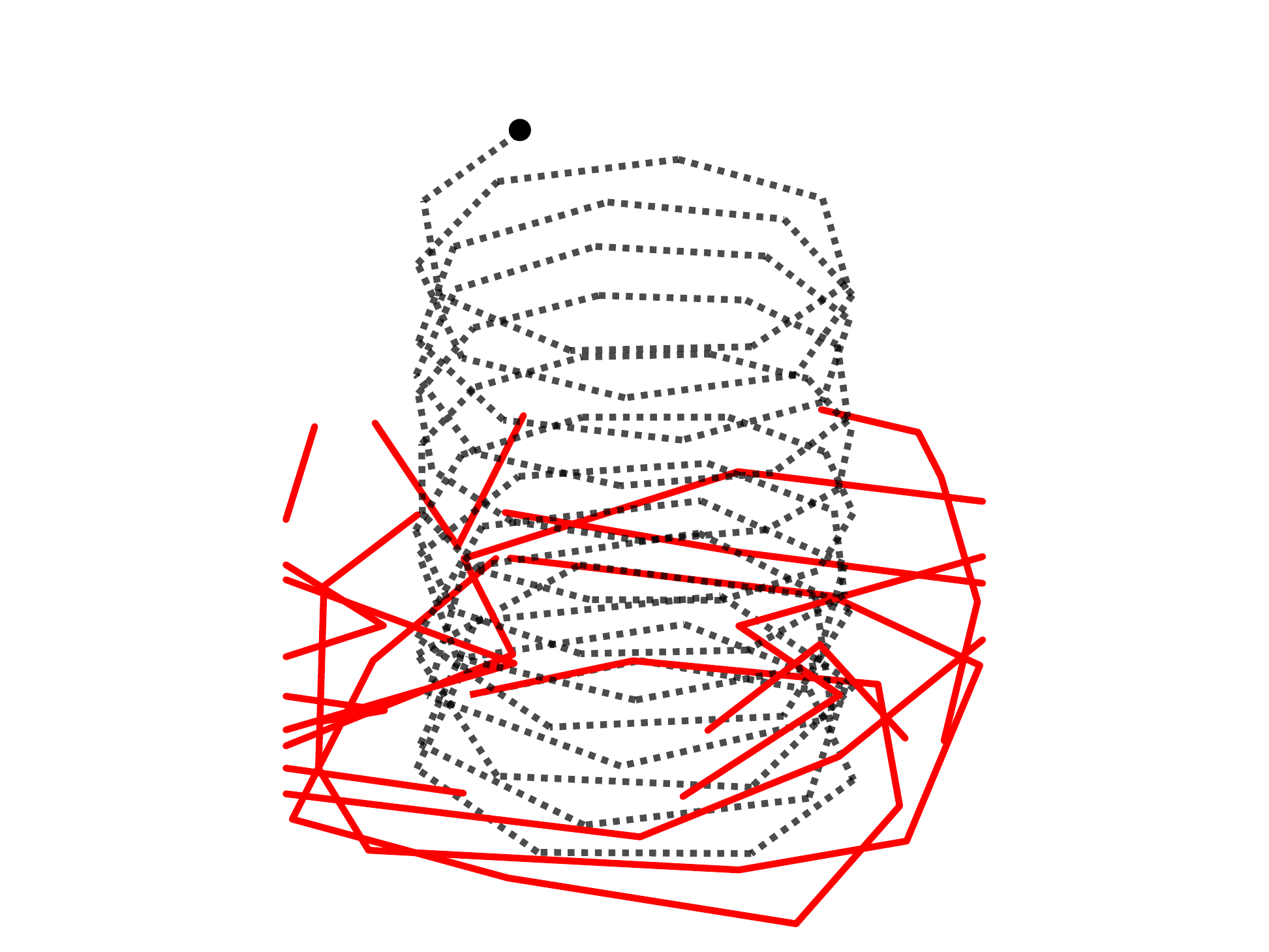}
			\caption{}
		\end{subfigure}
		\hspace{-1cm}
		\begin{subfigure}[b]{0.25\textwidth}
			\includegraphics[width=\linewidth]{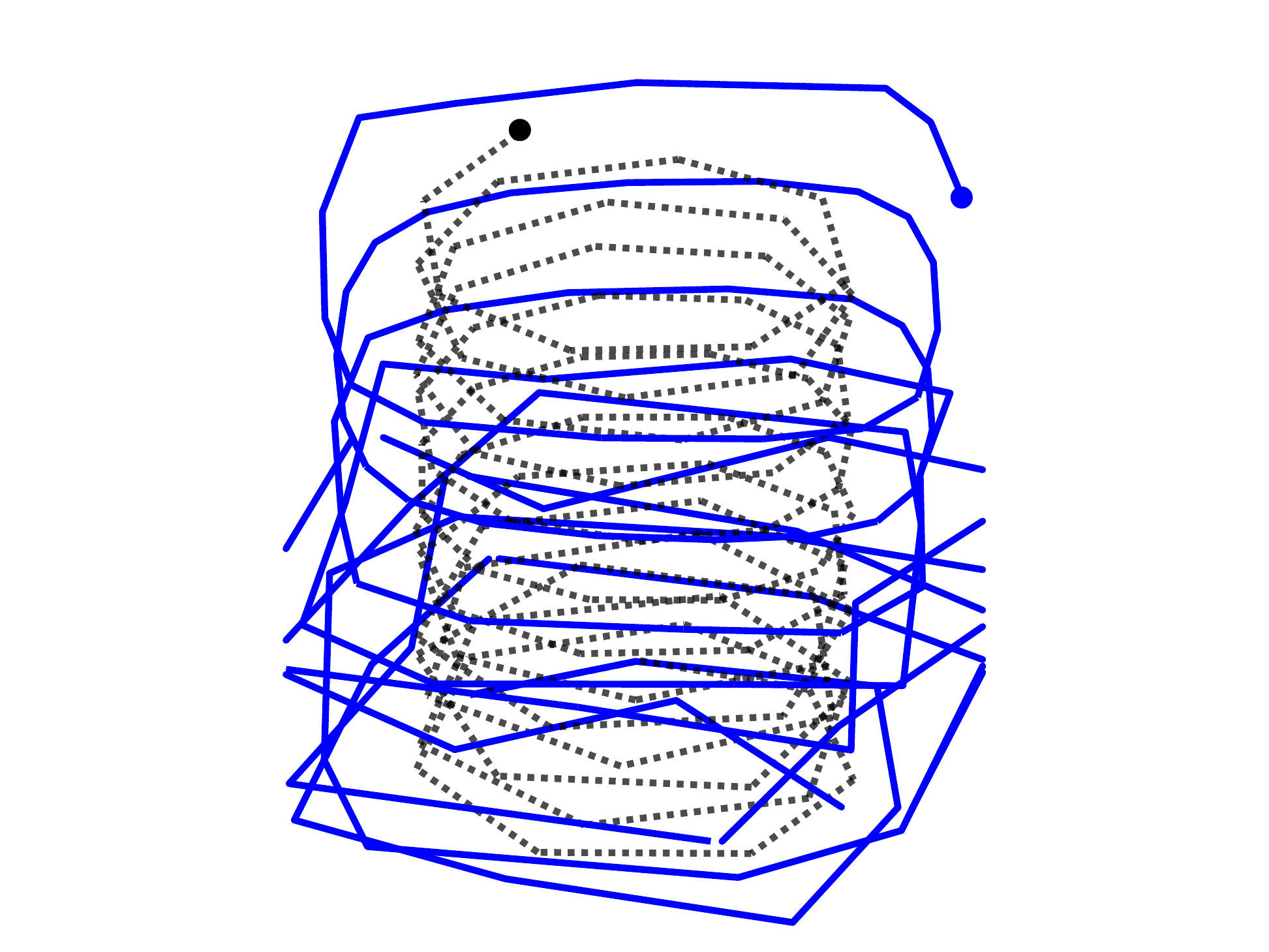}
			\caption{}
		\end{subfigure}
		\caption{Solution trajectories for MRBF+VP (left column, pink), MRBF+EMP (middle-left column, green), MRBF+AB2 (middle-right column, red) and TC+AB2 (right column, blue). The reference solution is given by a black dashed line. The rows correspond to time-steps of $h=\frac{1}{8},\frac{1}{4},\frac{1}{2}$ and $1$ from top to bottom. The grid size is $\Delta x = 1/2$ and the MRBF interpolant uses only $2\times 2\times 2$ data points. The MRBF interpolation uses inverse quadrics with shape parameter $\epsilon = 0.12$.}
		\label{TGVh}
	\end{figure}

\begin{figure}
		\centering
		\begin{subfigure}[b]{0.3\textwidth}
			\includegraphics[width=\linewidth]{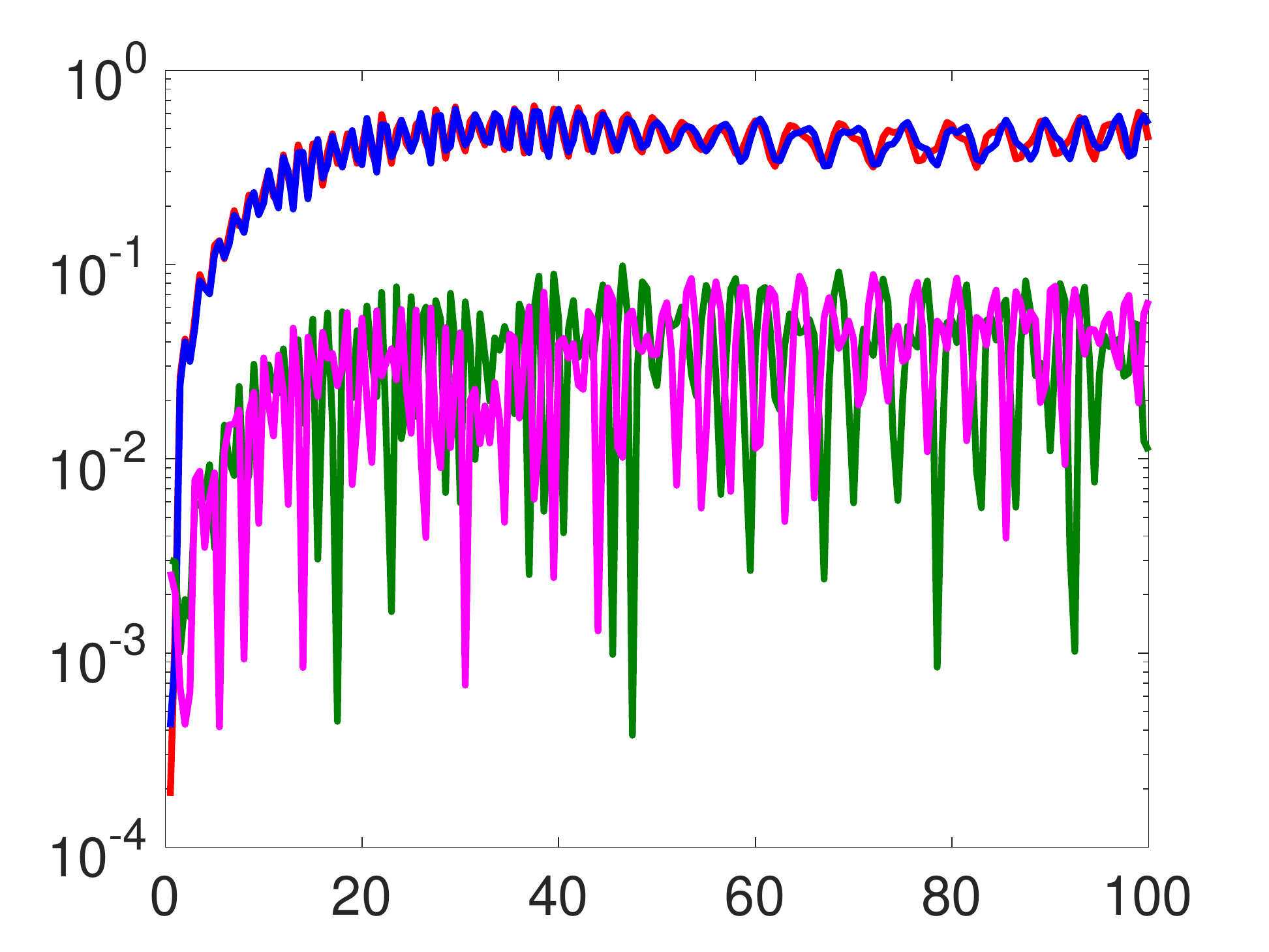}
			\caption{Radius error}
		\end{subfigure}
		\begin{subfigure}[b]{0.3\textwidth}
			\includegraphics[width=\linewidth]{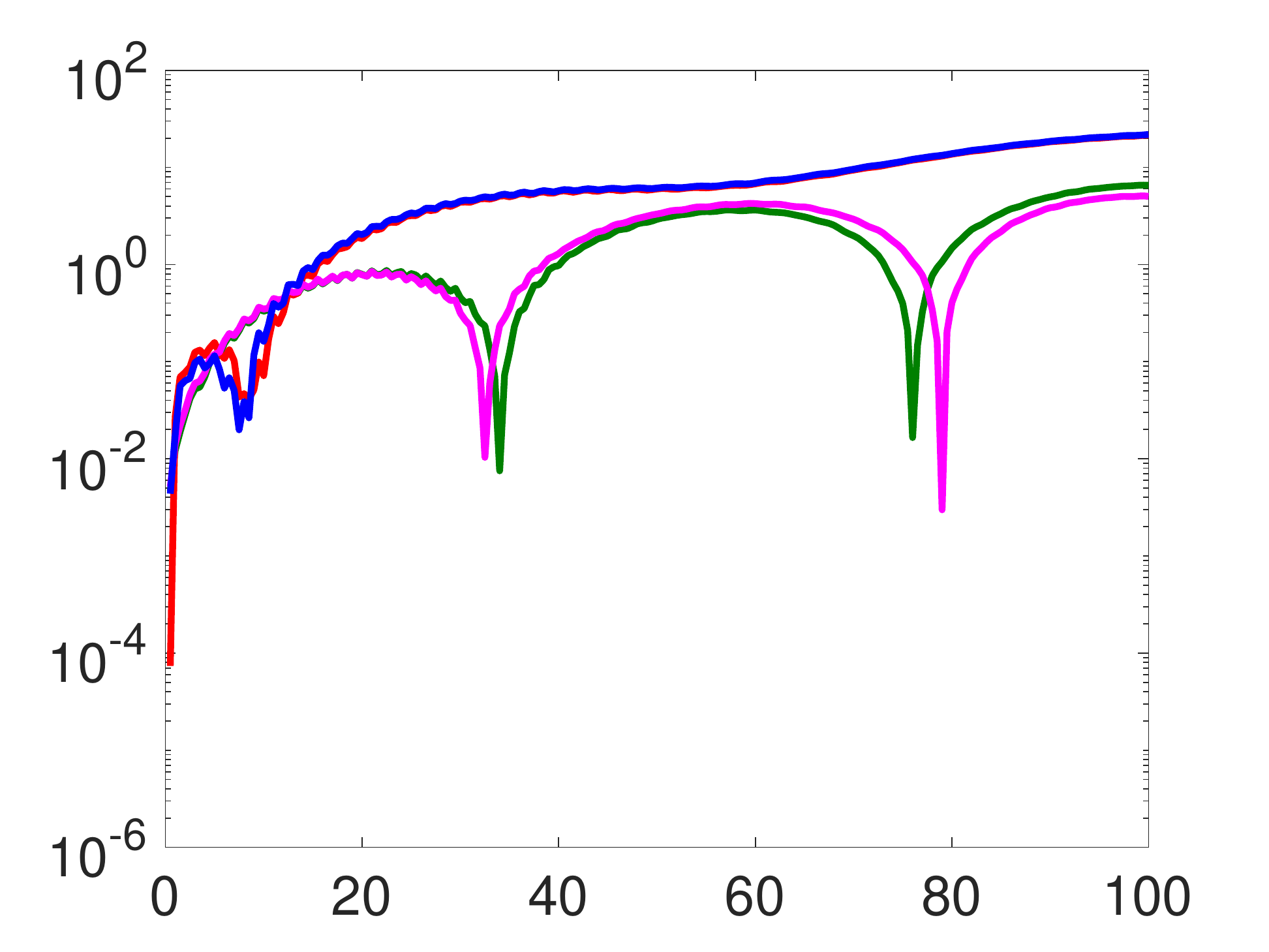}
			\caption{Phase error (rad)}
		\end{subfigure}
		\begin{subfigure}[b]{0.3\textwidth}
			\includegraphics[width=\linewidth]{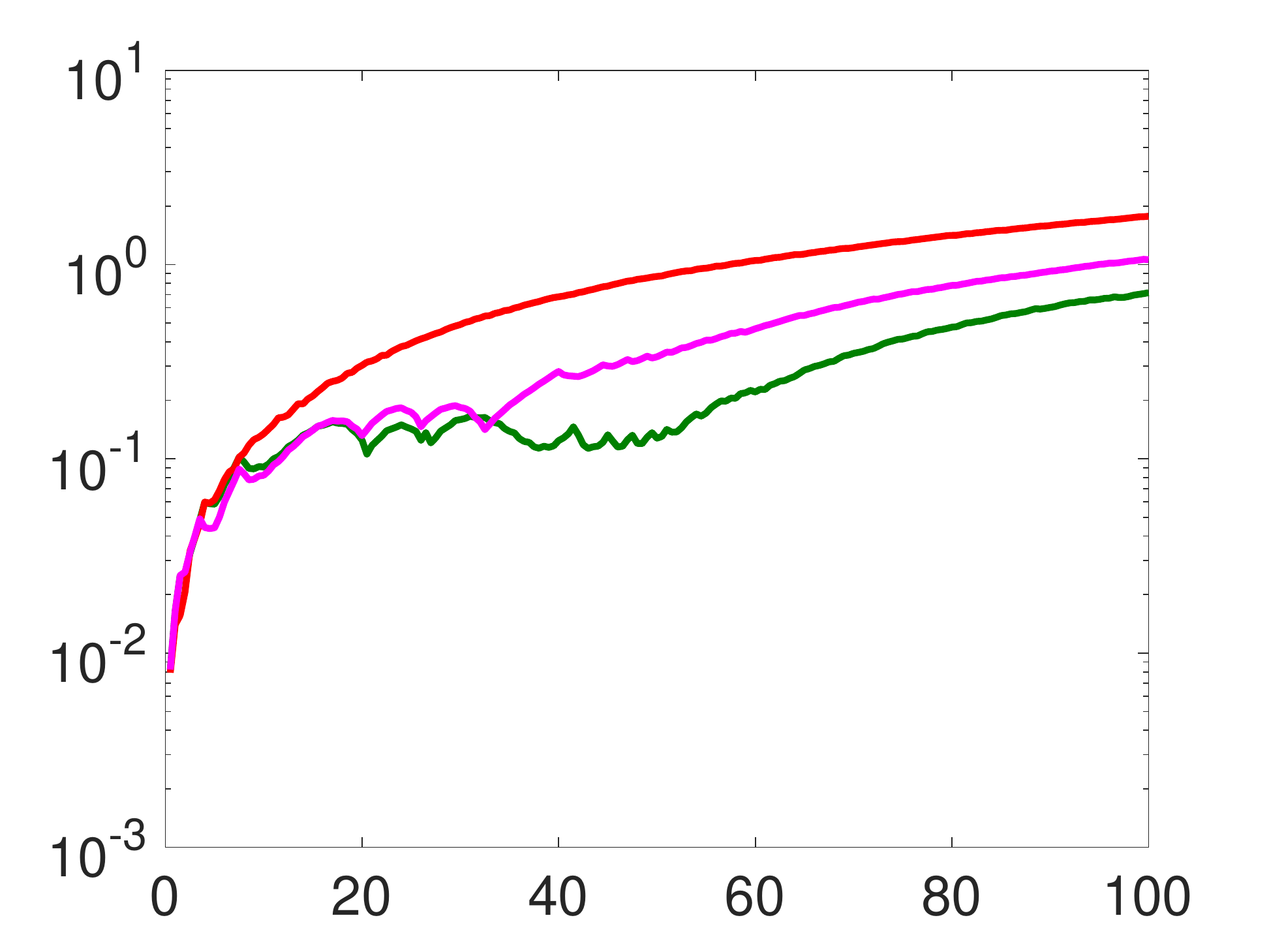}
			\caption{z-position (vertical) error}
		\end{subfigure}
		\hspace{3cm}
		\caption{The relative errors for the  $h = 1/2$ and $\Delta x = 1/2$ simulation.}
		\label{errors}
	\end{figure}

	Figure \ref{TGVh} shows the solution trajectories of the four methods for time-steps $h=\frac{1}{8},\frac{1}{4},\frac{1}{2}, 1$, from top to bottom. To emphasise the advantages of the MRBF schemes here, the interpolation now uses only the nearest $2\times2\times2$ data points for interpolation, which corresponds to solving a $32\times32$ linear system, which is roughly the same cost as a linear interpolation. The shape parameter is $\epsilon = 0.12$. Figure \ref{errors} shows the errors of the vortex radius, phase and z-position for the $h=\frac{1}{2}$ row in figure \ref*{TGVh}.\\
	
	There are a number of observations to be made in figure \ref{TGVh}. Perhaps the most outstanding one is that the green MRBF+EMP solution produces a remarkably accurate solution at $h=1$. For time-steps $h<1$, however both the magenta MRBF+VP and green MRBF+EMP methods produce vortices of radius very close to that of the reference solution. The main error in these two solutions are seen as phase errors and erroneous vertical velocities. We note that the red MRBF+AB and the blue TC+AB solutions look similar to the naked eye and are unable to capture the correct vortex dynamics for $h\ge\frac{1}{4}$. This supports the use of MRBF interpolation, inspite of an all-purpose integrator as it is comparatively cheaper compared to tricubic interpolation and can furnish similar trajectories at a fraction of the cost. Another noteworthy observation is that the end position of the particles do not converge to the true solution as $h$ is decreased. This is reflective of the fact that we are now in the saturation region where decreasing the time-step is of no benefit as the global error is polluted with interpolation error. The solution here is instead converging to the exact solution of the interpolated vector field. It is at this stage that we only see improvements in accuracy if one refines the interpolation methods. 
	
	\section{Conclusion}
	We show that using divergence-free radial basis functions for interpolating numerically calculated incompressible fluid fields can result in an efficient algorithm when combined with volume preserving maps for calculating pathlines. The resulting algorithm is implicit but we also suggest an explicit algorithm that exhibits the same qualitative features as the explicit algorithm in our particular numerical experiment. Compared to a conventional method, we show through numerical experiments that one can afford much greater step-size (about $4-8\times$ in our example) and use $56$ less data points for the interpolation step, whilst still providing the most long-term accurate solution. We also demonstrate that using divergence free interpolation is not enough to gain accurate trajectories for long-time simulations as stepping in time with a conventional method such as an Adams-Bashforth step will inevitably destroy the qualitative features of the solution and can lead to inaccurate particle trajectories. However, using RBF interpolation can still result in cheaper algorithms due to being able to capture reasonable divergence-free approximations to the fluid field with less interpolation points than a tricubic scheme.
	\section{Acknowledgements} 
	This work has received funding from the European Unions Horizon 2020 research and innovation
programme under the Marie Sklodowska-Curie grant agreement (No. 691070). 
	\section*{References}	
	\bibliography{bibliography}

\end{document}